\documentclass[%
letterpaper,
10pt,
accepted=2023-06-07,
]{quantumarticle}

\pdfoutput=1
\usepackage{graphicx}%
\usepackage{dcolumn}%
\usepackage{bm}%

\usepackage{amsmath,amssymb}
\usepackage[utf8x]{inputenc}
\usepackage{xcolor}
\usepackage{braket}
\usepackage{hyperref}
\usepackage[square,numbers]{natbib}
\usepackage{cleveref}
\usepackage{verbatim}
\usepackage[caption=false]{subfig}
\usepackage{graphicx}
\usepackage{multirow}
\usepackage{tikz}
\usepackage{lipsum}

\begin{document}

	\title{Classical models may be a better explanation of the Jiuzhang 1.0 Gaussian Boson Sampler than its targeted squeezed light model}
	
	\author{Javier Martínez-Cifuentes}
	\affiliation{%
		Department of Engineering Physics, \'Ecole Polytechnique de Montréal, Montréal, QC, H3T 1JK, Canada
	}%
	\author{K. M. Fonseca-Romero}
	\affiliation{  Departamento de Física, Universidad Nacional de Colombia - Sede Bogotá, Facultad de Ciencias, Grupo de Óptica e Información Cuántica, Carrera 30 Calle 45-03, C.P. 111321, Bogotá, Colombia}
	\author{Nicol\'as Quesada}
	\affiliation{%
		Department of Engineering Physics, \'Ecole Polytechnique de Montréal, Montréal, QC, H3T 1JK, Canada
	}%
	
	\begin{abstract}
		Recently, Zhong et al.~\cite{zhong2020quantum,zhong2021phase} performed landmark Gaussian boson sampling experiments with up to 144 modes using threshold detectors. The authors claim to have achieved quantum computational advantage with the implementation of these experiments, named Jiuzhang 1.0 and Jiuzhang 2.0.
		Their experimental results are validated against several classical hypotheses and adversaries
		using tests such as the comparison of statistical correlations between modes, Bayesian hypothesis testing and the Heavy Output Generation (HOG) test. In this work we propose an alternative classical hypothesis for the validation of these experiments. We use the probability distribution of mixtures of coherent states sent into a lossy interferometer; these input mixed states, which we term \textit{squashed states}, have vacuum fluctuations in one quadrature and excess fluctuations in the other.
		We find that for configurations in the high photon number density regime, the comparison of statistical correlations does not tell apart the ground truth of the experiment (namely, the probability distribution of two-mode squeezed states sent into a lossy interferometer) from our alternative hypothesis. On the other hand, the Bayesian test indicates that, for all configurations excepting Jiuzhang 1.0, the ground truth is a more likely explanation of the experimental data than our alternative hypothesis. A similar result is obtained for the HOG test: for all configurations of Jiuzhang 2.0, the test indicates that the experimental samples have higher ground truth probability than the samples obtained from our alternative distribution; for Jiuzhang 1.0 the test is inconclusive. Our results
		provide a new hypothesis that should be considered in the validation of future GBS experiments, and shed light into the need to identify proper metrics to verify quantum advantage in the context of threshold GBS. Additionally, they indicate that a classical explanation of the Jiuzhang 1.0 experiment, lacking any quantum features, has not been ruled out.
	\end{abstract}
	
	\maketitle
	
	\section{\label{sec:introduction}Introduction}
	One of the most exciting frontiers within the field of quantum computing is the topic of quantum advantage, which aims to design experiments that demonstrate the ability of current quantum devices to significantly outperform classical computers at a well defined computational task~\cite{harrow2017quantum,hangleiter2022computational}. The theoretical design and real-world implementation of such experiments has been the focus of significant effort in the superconducting circuit~\cite{arute2019quantum,wu2021strong} and quantum photonics~\cite{zhong2020quantum,zhong2021phase,madsen2022quantum} communities in the form of Random Circuit Sampling (RCS)~\cite{boixo2018characterizing,bouland2019complexity} and Gaussian Boson Sampling (GBS)~\cite{hamilton2017gaussian,kruse2019detailed,deshpande2022quantum,grier2021complexity}, respectively.
	
	GBS consists in sending a set of input squeezed states into an interferometer and measuring its output using photon-number~\cite{hamilton2017gaussian,kruse2019detailed} or threshold~\cite{quesada2018gaussian} detectors. The former  directly sample the photon number distribution while the later sample binary patterns of clicks that indicate whether light has been detected or not. It has been shown that sampling from the theoretical probability distribution of the resulting detection patterns, i.e., the ground truth distribution of the ideal experiment, is a computationally hard task~\cite{hamilton2017gaussian,deshpande2022quantum,grier2021complexity}.
	These results have important caveats, for example, the known proofs of hardness for GBS require the photon number density (the average number of photons per mode) to be small so that the probability of two or more photons being measured in the same detector is very small.
	On the algorithmic side, the best known methods to classically simulate these quantum sampling problems scale exponentially in the number of detected photons or counted clicks~\cite{bulmer2022boundary,quesada2022quadratic,quesada2020exact,gupt2020classical,bourassa2021fast,chabaud2022resources}.

	Verifying that quantum samplers are operating correctly remains an active area of research (cf. the review paper by Hangleiter and Eisert~\cite{hangleiter2022computational} and references therein). For RCS it was identified early on that an estimate of the cross entropy between the samples generated by the physical device and the probability distribution associated with the ideal computation served as a witness of quantum advantage~\cite{boixo2018characterizing,bouland2019complexity}.
	For GBS the situation is more complex, as the development of a proper figure of merit that allows to readily verify a claim of quantum computational advantage is still an open challenge~\cite{hangleiter2022computational,deshpande2022quantum}.
	On this account, the validation of GBS usually relies on a series of tests that rule out possible classical hypotheses or that compare the quality of the samples generated by the quantum machine against samples generated by classically efficient methods.
	
	Two recent landmark threshold GBS experiments by Zhong et al. using 100-mode~~\cite{zhong2020quantum} and 144-mode~\cite{zhong2021phase} interferometers (named Jiuzhang 1.0 and 2.0, respectively) claim to achieve quantum computational advantage.
	The authors use three different validation tests: the comparison of truncated first to fourth order correlation functions (i.e., the first to fourth order click cumulants of the probability distribution), a Bayesian test and the Heavy Output Generation (HOG) test.
	The first examines how well the correlations in the observed data match the correlations predicted by the squeezed state hypothesis.
	The second test compares how good the squeezed state hypothesis is at explaining the observed data relative to other hypotheses such as thermal states, coherent states, distinguishable squeezed states and uniform probability distributions.
	The third test looks at how well the samples generated by the experiment have ``heavy outputs'' (i.e., correspond to events with high probability) in the ideal distribution relative to samples generated by classically efficient methods. These classically efficient methods can be physically motivated as considered by Zhong et al.~\cite{zhong2020quantum,zhong2021phase} but need not be~\cite{villalonga2021efficient}.

	These validation tests intend to build up confidence in the correct functioning of the boson sampler. However, they have several limitations and by no means represent a complete measure to readily tell if a device achieves quantum advantage. For instance, the Bayesian and HOG tests rely on the computation of probabilities of a great number of detection patterns, a task that increases its complexity exponentially with the number of detected photons (clicks) present in the patterns. Consequently, they can only be used in a small region of the photon (click) number distribution.  
	
	In this work, we propose an alternative classical hypothesis for the validation of the experiments of Zhong et al. Our hypothesis is based on the probability distribution that is obtained from using classical mixtures of coherent states (resulting in Gaussian states that we term \emph{squashed states}~\cite{qi2020regimes,jahangiri2020point}) as inputs of the interferometers in GBS setups. These states are classical, possessing a non-negative Glauber-Sudarshan $P$ function~\cite{reid1986violations,drummond2014quantum,rahimi2016sufficient,rahimi2015can} and upon interacting on an interferometer generate fully separable (i.e. having zero entanglement) multimode states (by virtue of having a positive multimode Glauber-Sudarshan $P$ function).

	To compare the squeezed states and squashed states hypotheses, we investigate two of the tests (correlation functions and Bayesian test) that the authors of the experiment employed.
	We find that for configurations in the high photon number density regime (which correspond to the Jiuzhang 1.0 experiment and the two brightest configurations of Jiuzhang 2.0), on which most of the quantum computational advantage claim relies, the truncated correlation functions predicted by the squashed states distributions are as consistent with the experimental results as those predicted by the ground truth of the experiment given by pure squeezed states. On the other hand, the Bayesian test shows that the ground truth of the experiment is more likely to describe the experimental samples of Jiuzhang 2.0 than the squashed states hypothesis. For Jiuzhang 1.0, the test indicates that the squashed states hypothesis is more likely. This behaviour of the Bayesian test is consistent with the improvement of the light source of the Jiuzhang 2.0 experiment with respect to Jiuzhang 1.0.

	These results demonstrate that, contrary to what the authors of the experiment suggested, the comparison of correlation functions is not sufficiently reliable for the verification of quantum advantage claims. Additionally, they indicate that the squashed states hypothesis presents itself as viable classical hypothesis for the validation of current threshold GBS experiments that should be weighted in future experiments.
	
	After considering hypothesis testing, we generate samples from our squashed states hypothesis (which can be done in polynomial time and space in the number of modes/clicks~\cite{rahimi2016sufficient,qi2020regimes,gupt2019thewalrus}) and compare against the experimental samples from the Jiuzhang experiments, to see if, like the Boltzmann machine and greedy methods from Villalonga et al.~\cite{villalonga2021efficient}, we can spoof the HOG test. This test compares the probabilities of the experimental samples and the samples from the squashed states with respect to the ground truth. Surprisingly, for the Jiuzhang 1.0 experiment, the test is inconclusive.	
	For the Jiuzhang 2.0 experiment, on the other hand, we find that the squashed states samples are unable to spoof the HOG test. 
	
	Besides our analysis of the experiments of Zhong et al., there have been other works that investigate classical ways to simulate some properties of the experimental data of the Jiuzhang experiments, or that directly aim to produce classical samples that perform better than the experimental patterns at different statistical tests. The first of them, by Drummond et al.~\cite{drummond2022simulating}, proposes a method to classically simulate grouped click probability distributions of general GBS setups. The authors use this method to simulate different grouped click distributions of the Jiuzhang 1.0 experimental data. Moreover, they propose a model that combines input squeezed thermal states and a modification of the intensity transmissivity of the interferometer, in order to take into account several experimental sources of decoherence, such as phase noise or the presence of longitudinal modes with mismatches in time or frequency. The authors find that their decoherence model has a better fit to the experimental results than the ground truth of the experiment. The second of these works, by Villalonga et al.~\cite{villalonga2021efficient}, proposes a classical algorithm, based on marginal distributions of the ground truth, to generate samples that have better total variation distance and Kullback-Leibler divergence with respect to the ground truth than the experimental samples of the Jiuzhang experiments. Together with our results, these works suggest that a classical explanation of the experiments of Zhong et al. has not been completely ruled out.
	
	This paper is organized as follows: In Sec.~\ref{sec:distributions} we obtain the ground truth distribution of the Zhong et al. experiments and define the squashed states hypothesis. In Sec.~\ref{sec:validation} we specify the different tests used in the validation of the experimental samples against the squashed states hypothesis and show the corresponding results. Finally, we conclude in Sec.~\ref{sec:discussion}.

	\section{\label{sec:distributions}Ground truth and squashed states distributions}
	
	\begin{figure*}[!ht]
		\centering
		\subfloat[]{\label{fig:scheme}
			\includegraphics[scale=0.39]{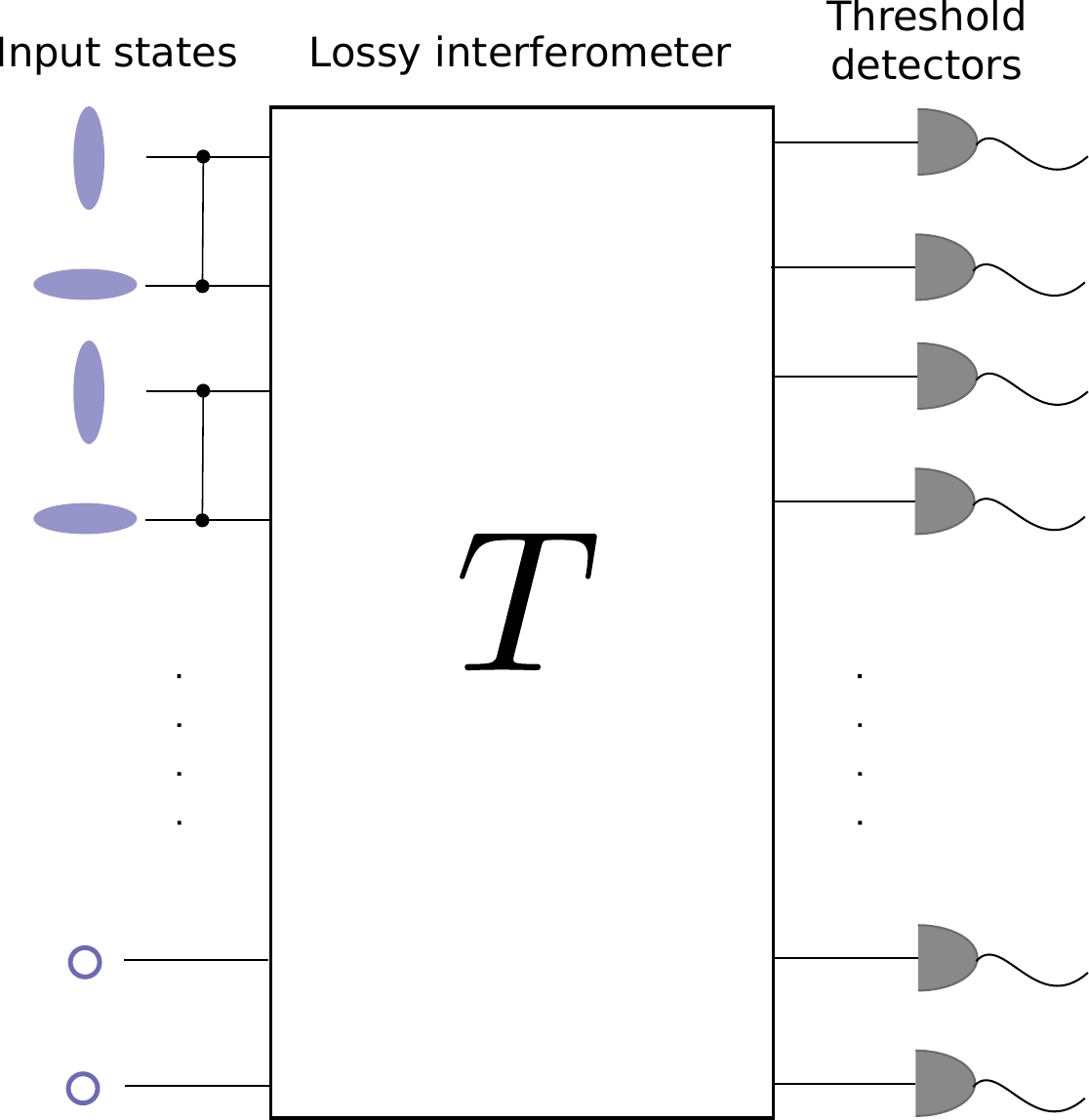}
		}
		\hfill
		\subfloat[]{\label{fig:hypotheses}
			\includegraphics[scale=0.39]{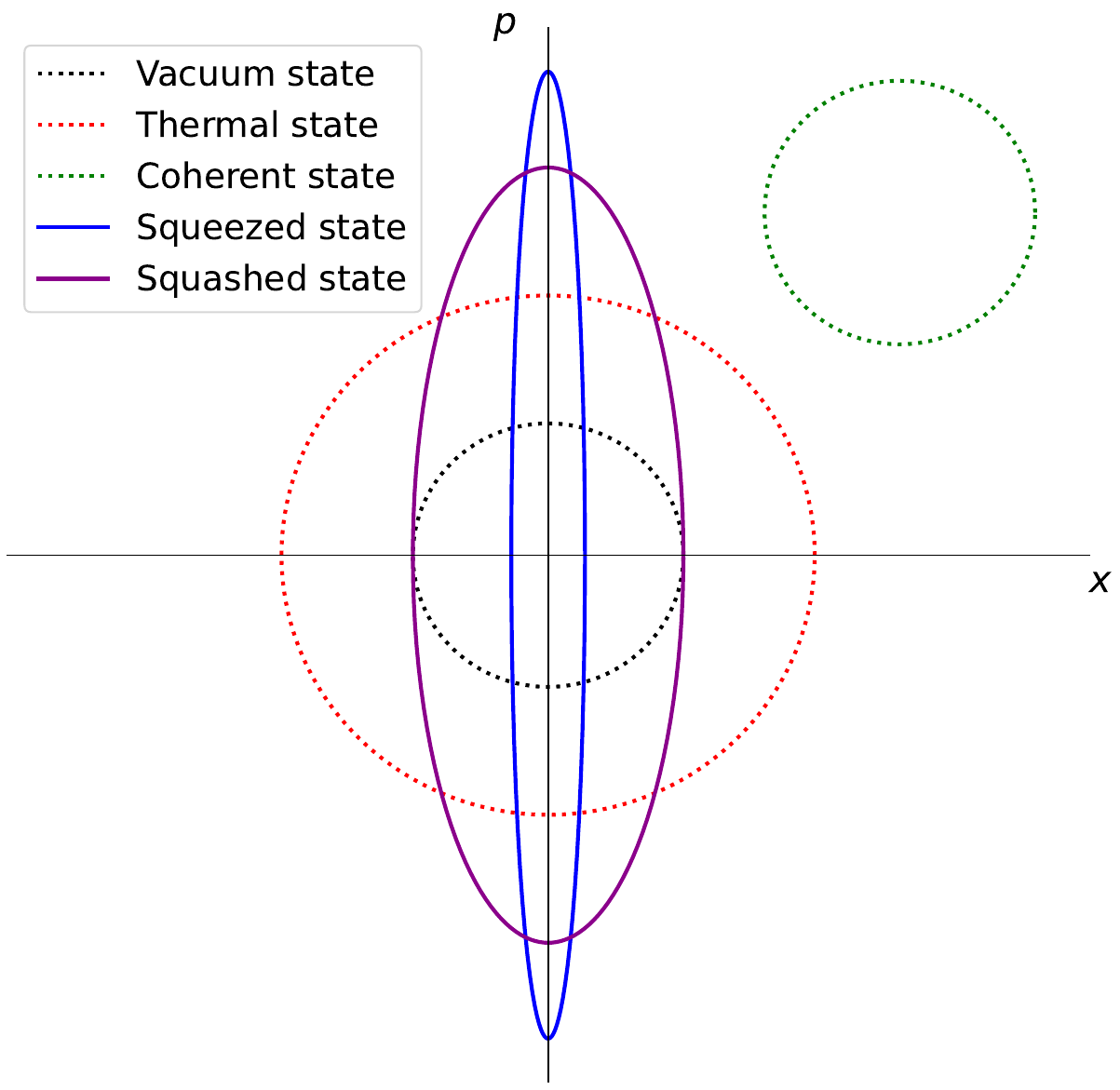}
		}
		\caption{(a) Description of the Jiuzhang 1.0/2.0 GBS hypotheses. Pairs of adjacent input squeezed or squashed states are sent into 50:50 beamsplitters represented by the vertical lines with dots at the end. The alternating orientation of the input states together with the 50:50 beamsplitters generate pairs of two-mode squeezed or squashed states. These states are then sent into a lossy interferometer, represented by the rectangle, whose action is described by a rectangular sub-unitary matrix $\bm{T}$. The output state of the interferometer is sampled using threshold detectors. (b) Representation of the noise ellipse of several Gaussian states used for the validation of the Jiuzhang 1.0/2.0 experiments. The ground truth of the experiments corresponds to the use of squeezed states corresponding to the blue line. Jiuzhang 1.0 and Jiuzhang 2.0 have been validated against hypotheses using coherent states and thermal states represented by the red-dotted and green-dotted lines, respectively. In this work we validate the experimental results against a hypothesis using the squashed states represented by the purple line. For reference, we also show the ellipse corresponding to the vacuum state with a black-dotted line.}
		\label{fig:experiment}
	\end{figure*}
	
	A threshold GBS experiment has three stages: preparation of $K$ (single-mode) squeezed states, evolution in an interferometer with $M\geq K$ output modes, and sampling of the output statistics using threshold detectors.
	These detectors do not resolve the incoming number of photons, they can only indicate if light has arrived to the detector or not.
	Therefore, the outcomes of the measurement can be expressed as $M$-bit strings, patterns consisting only of zeros and ones, where one indicates that a detector has been triggered (the detector has `clicked') and zero indicates that no light has been detected.

	Jiuzhang 1.0 (2.0) uses 25 two-mode squeezed states (TMSS), corresponding to 50 squeezed states, as inputs to a 100-mode (144-mode) interferometer for the first (second) setup. According to Zhong et al.~\cite{zhong2021phase}, the light source of Jiuzhang 2.0 is greatly improved with respect to that of Jiuzhang 1.0, in the sense that its quantum state is closer to the theoretical model (an affirmation that is supported by recent theoretical studies~\cite{houde2022waveguided}), and allows the implementation of seven different configurations by varying its power and focus waist.
	
	\begin{table*}[t!]
		\centering
		\begin{tabular}{|c|c|c|c|c|c|}
			\hline
			
			\rule{0pt}{11pt} Experiment & $P(\text{W})$ & $w(\mu\text{m})$ & $\nu$ & $\bar{C}$ & $\sigma(C)$\\ \hline
			Jiuzhang 1.0& - & - & $0.786$ & $41.042$ & $6.509$\\ \hline
			\multirow{7}{*}{Jiuzhang 2.0}
			& $0.5$ & $125$ & $0.055$ & $7.273$ & $3.391$\\ \cline{2-6}
			& $1.412$ & $125$ & $0.161$ & $19.256$ & $5.596$\\ \cline{2-6}
			& $0.15$ & $65$ & $0.044$ & $5.976$ & $3.016$\\ \cline{2-6}
			& $0.3$ & $65$ & $0.093$ & $11.941$ & $4.325$\\ \cline{2-6}
			& $0.6$ & $65$ & $0.218$ & $24.664$ & $6.269$\\ \cline{2-6}
			& $1.0$ & $65$ & $0.442$ & $41.794$ & $7.95$\\ \cline{2-6}
			& $1.65$ & $65$ & $0.975$ & $66.866$ & $9.02$\\ \hline
		\end{tabular}
		\caption{Available specifications (power $P$ and focus waist $w$) of the light source and theoretical photon density $\nu=\frac{\bar{N}}{M} = \frac{1}{M} \sum_{i=1}^M \langle a_i^\dagger a_i \rangle$, theoretical mean number of clicks $\bar{C}$, and theoretical standard deviation of the number of clicks $\sigma(C)$ of the different configurations of Jiuzhang 1.0 (with $M=100$ modes) and Jiuzhang 2.0 experiments (with $M=144$ modes).
		}
		\label{tab:experiment_specifics}
	\end{table*}
	
	The specification of the different experimental configurations, along with the theoretical photon number density, $\nu = \bar{N} / M$ (the ratio between the mean photon number $\bar{N}$ and the number of output modes), mean number of clicks, $\bar{C}$, and standard deviation of the number of clicks, $\sigma(C)$, of the Jiuzhang 1.0 and Jiuzhang 2.0 experiments are shown in Table~\ref{tab:experiment_specifics}. The values of $\bar{C}$ and $\sigma(C)$ were computed using the methods of Refs.~\cite{grier2021complexity, gupt2019thewalrus}.

	The ground truth of the Jiuzhang 1.0 and Jiuzhang 2.0 experiments is described as 25 input two-mode squeezed states that interfere in a linear lossy interferometer. The output state is then measured using ideal threshold detectors. The interferometer in this model contains the information about the losses in the real experiment (collection efficiency, propagation loss, and finite detector efficiency).
	
	Two-mode squeezed states are mathematically defined as
	\begin{equation}
		|\zeta_{kl}\rangle=\exp\left(-\zeta a_k^\dagger a_l^\dagger+\zeta ^{*} a_l a_k\right)|0_k,0_l\rangle,
		\label{eq:TMSS_def}    
	\end{equation}
	where $|0_k,0_l\rangle$ is the vacuum state of modes $k$ and $l$, $a_k, a_l$ and $a_k^\dagger, a_l^\dagger$ are the annihilation and creation operators of these modes, and $\zeta=re^{i\phi}$ is a complex squeezing parameter. The annihilation and creation operators of the modes satisfy the usual bosonic canonical commutation relations $[a_i,a_j] = [a_i^\dagger, a_j^\dagger] = 0 $ and  $[a_i, a_j^\dagger] = \delta_{i,j}$.
	The state $|\zeta_{kl}\rangle$ can also be expressed in the form~\cite{serafini2017quantum, barnett2002methods}
	\begin{equation}
		|\zeta_{kl}\rangle=U_\text{BS}\left[S_k(\zeta)|0_k\rangle\otimes S_l(-\zeta)|0_l\rangle\right],
		\label{eq:TMSS_to_SMSS}
	\end{equation}
	where $U_\text{BS} = \exp[\frac{\pi}{4}(a_l^\dagger a_k - a_k^\dagger a_l)]$ is a unitary transformation representing the action of a $50:50$ real beamsplitter\footnote{We use the term real beamsplitter to indicate that the coefficients relating $a_l$ ($a_l^\dagger$) and $a_k$ ($a_k^\dagger$) in $U_\text{BS}$ are all real.}, and $S_k(\zeta)=\exp[-\frac{1}{2}(\zeta a_k^{\dagger2} - \zeta^* a_k^2)]$ is the single-mode squeezing operator of mode $k$. This expression indicates that a TMSS can be physically generated by interfering two single-mode squeezed states (SMSS) compressed along orthogonal directions into a $50:50$ real beamsplitter.
	
	For Jiuzhang 1.0 and Jiuzhang 2.0, the squeezing parameters of the input states $\zeta_i=r_i,$ $i=1,\cdots,25,$ are considered to be real and positive.
	The information about the phases $\phi_i$ is absorbed into the rectangular sub-unitary matrix $\bm{T}$ that describes the action of the lossy interferometer. As shown in Eqs.~\eqref{eq:sigma_out} and~\eqref{eq:vmat}, this matrix allows the calculation of the output state for Gaussian input states.	The size of $\bm{T}$ is $M \times 50$, where $M$ is the number of output modes.
	Data about the squeezing parameters and the matrix $\bm{T}$ is available in~\cite{ustc2020experimental} for Jiuzhang 1.0, and in~\cite{ustc2021raw} for Jiuzhang 2.0.
	
	In order to verify the results of the Jiuzhang experiments, we need to determine the theoretical probability distribution of the experimental samples, which is commonly known as the \textit{ground truth distribution} of the experiment. The specification of this distribution will also allow us to motivate the definition of the probability distribution of the squashed states.
	
	The ground truth distributions of the Jiuzhang experiments are completely characterized by the output state of the interferometer.
	Since the input states are Gaussian, and the interferometer is linear, the output state is also Gaussian.
	Gaussian states of $K$ modes are characterized by their vector of first moments and their covariance matrix.
	If we write the annihilation operators as $a_k = (x_k + i p_k)/\sqrt{2\hbar},$ in terms of the hermitian in- (out-~) phase quadratures $x_k$ $(p_k)$, we can define the vector $\bm{r}^\text{T} = (x_1, x_2, \cdots, x_K,p_1, p_2, \cdots, p_K)$. Then, assuming a vanishing vector of first moments $\overline{\bm{r}} = \operatorname{Tr} (\rho_G \bm{r}) = \bm{0},$ (as is the case for the Jiuzhang experiments) the covariance matrix can be written as $\bm{\sigma} = \frac{1}{2}\operatorname{Tr} \left(\rho_G \{ \bm{r}, \bm{r}^T\}\right).$
	Here, $\rho_G$ denotes the density matrix of the Gaussian state and $\{A,B\} = AB + BA$ stands for the anticommutator of operators $A$ and $B$. On this account, the ground truth distribution of the Jiuzhang experiments will be completely specified by the covariance matrix of the output states. In what follows, we will explain how to compute this matrix.
	
	We first specify the covariance matrix of the input states. To do this, we follow the procedure indicated by Eq.~\ref{eq:TMSS_to_SMSS}: we generate 25 TMSS by interfering 50 pairs of SMSS with squeezing parameters $\{-r_1, r_1,\dots,-r_{25}, r_{25}\}$ into 25 beamsplitters that act on consecutive modes. The covariance matrix of one SMSS with squeezing parameter $r$ is
	\begin{equation}
		\bm{\sigma}_\text{SMSS}^{(1)}=\frac{\hbar}{2}
		\begin{pmatrix}
			e^{-2r}&0\\
			0&e^{2r}
		\end{pmatrix}.
		\label{eq:one_SMSS}    
	\end{equation}
	Note that in this covariance matrix the variances of the two quadratures are inversely related, thus saturating the uncertainty relation~\cite{serafini2017quantum}.
	
	We can now construct the $100\times100$ covariance matrix of the 50 SMSS (in the ordering indicated by $\bm{r}^\text{T}$) as
	\begin{align}
		\begin{split}
			\bm{\sigma}_\text{SMSS} = \frac{\hbar}{2}\,& \text{diag}\left(e^{2r_1},\, e^{-2r_1},\dots,\,e^{2r_{25}},\, e^{-2r_{25}},\right.\\
			&\left.\,e^{-2r_1},\, e^{2r_1},\dots,\,e^{-2r_{25}},\, e^{2r_{25}}\right).
		\end{split}
		\label{eq:single_squeezed_cov}
	\end{align}
	where $\text{diag}(v_1,\ldots,v_m)$ forms a diagonal matrix of size $m$ with entries $v_1,\ldots,v_m$.
	A $50:50$ real beamsplitter interfering two adjacent modes (see Fig.~\ref{fig:scheme}) is represented by the matrix
	\begin{equation}
		\bm{H}=
		\frac{1}{\sqrt{2}}
		\begin{pmatrix}
			1 & -1 \\
			1 & 1
		\end{pmatrix}.
		\label{eq:beam_splitter}    
	\end{equation}
	The corresponding matrix for the 25 beamsplitters has the form
	\begin{equation}
		\bm{B}=\bigoplus_{k=1}^{50}\bm{H},
		\label{eq:beam_splitters}    
	\end{equation}
	The covariance matrix of the input TMSS can then be computed as
	\begin{equation}
		\bm{\sigma}_\text{TMSS} = \bm{B}\bm{\sigma}_\text{SMSS}\bm{B}^\text{T}.
		\label{eq:input_TMSS_cov}
	\end{equation}
	
	Given a $K$-mode input Gaussian state with covariance matrix $\bm{\sigma}_{\text{IN}}$, an interferometer channel with transmission matrix $\bm{T}$ maps it to an $M$-mode Gaussian state with covariance matrix $\bm{\sigma}_{\text{OUT}}$  given by~\cite{serafini2017quantum}
	\begin{align}
	\begin{split}
		\bm{\sigma}_{\text{OUT}} &= \mathcal{L}_{\bm{T}}(\bm{\sigma}_{\text{IN}})\\
		&=\frac{\hbar}{2}\left(\mathbb{I}_{2M} - \bm{V}\bm{V}^{T}\right) + \bm{V}\bm{\sigma}_\text{IN}\bm{V}^\text{T},
	\end{split}
    \label{eq:sigma_out}
	\end{align}
	where
	\begin{align}	
		\bm{V}&=
		\begin{pmatrix}
			\text{Re}(\bm{T}) & -\text{Im}(\bm{T})\\
			\text{Im}(\bm{T}) & \text{Re}(\bm{T})
		\end{pmatrix}.
    \label{eq:vmat}
	\end{align}
	The complex transmission matrix in the last equation is generally rectangular and of dimensions $M \times K$.
	With this notation in mind, we can then simply write the covariance matrix of the squeezed ground truth hypothesis as
	$    \bm{\sigma}_{(\text{SQUE})} = \mathcal{\bm{L}}_{\bm{T}}(\bm{\sigma}_{\text{TMSS}})$.
	Having the covariance matrix $\bm{\sigma}$ of a given Gaussian state we can calculate click-probabilities as~\cite{quesada2018gaussian}
	\begin{align}
	\label{eq:ground_truth_distribution}
		\mathrm{Pr}(\bm{s})= \frac{\text{Tor}\left[\bm{O}_{(\bm{s})}\right]}{\sqrt{\text{det}(\bm{\Sigma})}},
	\end{align}
	with  $\bm{O} = \mathbb{I}_{2M} - \bm{\Sigma}^{-1}$, $\bm{\Sigma} = \frac{1}{2}\mathbb{I}_{2M} + \frac{1}{\hbar} \bm{R} \bm{\sigma} \bm{R}^\dagger$, and
	\begin{align}
		\label{eq:ground_truth_distribution_2}
		\bm{R} = \frac{1}{\sqrt{2}}
		\begin{pmatrix}
			\mathbb{I}_M & i\mathbb{I}_M\\
			\mathbb{I}_M & -i\mathbb{I}_M
		\end{pmatrix}.
	\end{align}
	Here,  $\text{Tor}(\bm{A})$ is the Torontonian of matrix $\bm{A}$ (see Appendix~\ref{app:tor} for a definition). The operation $\bm{A}_{(\bm{s})}$ is explained as follows: suppose that the detection pattern $\bm{s}$ contains $C$ clicks ($C$ ones) observed in the modes $\{j_1,\dots,j_C\}$, then matrix $\bm{A}_{(\bm{s})}$ is obtained by keeping only the rows and columns $\{j_1,\dots,j_C,j_1 + M,\dots,j_C + M\}$ of matrix $\bm{A}$, whose size is $2M \times 2M$.
	To obtain probabilities associated with the ground truth we simply let $\bm{\sigma} \to \bm{\sigma}_{(\text{SQUE})}$ in the last equation. Note that the results presented here can be extended to Gaussian states with displacements~\cite{thekkadath2022experimental} using loop Torontonians~\cite{bulmer2022threshold}.
	
	We now turn to the definition of the squashed states distribution. When losses are incorporated in the input states, squeezed states become squeezed thermal states~\cite{qi2020regimes}.
	These states still have a quadrature with noise lower than that of the vacuum state. We define the squashed states as squeezed thermal states with vacuum fluctuations in one quadrature and larger fluctuations in the other (in Appendix~\ref{app:squashed} we list some properties of these states and make a comparison of their covariance matrix with that of squeezed thermal states). The output states obtained by interfering squashed states are classical, which means that we can efficiently sample from their probability distribution using classical computers. As can be seen in Fig.~\ref{fig:hypotheses}, the noise ellipse of the squashed states suggests that they are better approximations to the squeezed states than the classical states that Zhong et al. used for the validation of the experimental results, which have circular noise ellipses. Indeed, squashed states are the classical Gaussian states with the highest fidelity to squeezed thermal states~\cite{qi2020regimes}.
	
	The definition of the probability distribution associated with the squashed states is now straightforward: we need to apply the interferometer channel with transmission $\bm{T}$ to 25 two-mode squashed states. In turn, the input covariance matrix to the interferometer is obtained by replacing $\bm{\sigma}_\text{TMSS}$ for the input 25 two-mode squashed states covariance matrix, $\bm{\sigma}_0'$. This matrix is obtained in the same way we constructed $\bm{\sigma}_\text{TMSS}$: we interfere 50 single-mode squashed states into 25 real $50:50$ beamsplitters. The covariance matrix of a single-mode squashed state with mean photon number $\bar{n}=\sinh^2r$ is
	\begin{equation}
		\bm{\sigma}_\text{0}^{(1)}=\frac{\hbar}{2}
		\begin{pmatrix}
			1&0\\
			0&1+4\bar{n}
		\end{pmatrix}.
		\label{eq:one_squashed}    
	\end{equation}
	This covariance matrix has no squeezing as one of the quadratures is at the vacuum level while the other has excess (classical) noise proportional to $\bar{n}$, thus they can be described as classical mixtures of coherent states as shown in Ref.~\cite{jahangiri2020point}.
	Notice that the matrix above can be obtained by replacing $e^{-2r}$ and $e^{2r}$ in Eq.~\eqref{eq:one_SMSS} by 1 and $1+4\bar{n}$, respectively. This same procedure can be used to construct the covariance matrix of the 50 single-mode squashed states, $\bm{\sigma}_0$: we replace the $e^{-2r_k}$ and $e^{2r_k}$ terms in Eq.~\eqref{eq:single_squeezed_cov} by 1 and $1+4\bar{n}_k$, with $\bar{n}_k=\sinh^2r_k$, respectively. Then, $\bm{\sigma}_0'$ can be written as
	\begin{equation}
		\bm{\sigma}_0' = \bm{B}\bm{\sigma}_0\bm{B}^\text{T}.
		\label{eq:input_squashed_cov}
	\end{equation}
	
	We can write the covariance matrix associated with the squashed states hypothesis as $\bm{\sigma}_{\text{(SQUA)}} = \mathcal{L}_{\bm{T}}(\bm{\sigma_0}')$ .
	The covariance matrix of this state satisfies $\bm{\sigma}_{\text{(SQUA)}} \ge \frac{\hbar}{2} \mathbb{I}_{2M}$ and thus the Gaussian state associated with it can be written as a mixture of products of single-mode coherent states~\cite{jahangiri2020point,rahimi2015can,rahimi2016sufficient}, implying that this Gaussian state is separable across any partition of its modes.
	Finally, having the covariance matrix of the state, we can obtain probabilities associated with it by letting $\bm{\sigma} \to \bm{\sigma}_{(\text{SQUA})}$ in Eq.~\eqref{eq:ground_truth_distribution}.

    It is worth mentioning that, as was proposed in Ref.~\cite{drummond2022simulating}, the transmission matrix of the interferometer, as well as the mean photon numbers of the input squashed states, can be modified in order to obtain a model that takes into account other sources of experimental decoherence (not only losses in the input squeezed states). However, we choose the squashed states to have the same mean photon numbers as the squeezed states used in the definition of the covariance matrix in Eq.~\eqref{eq:input_TMSS_cov} so as to have a photon number distribution with lower total variation distance to the distribution predicted by the ground truth. Moreover, we maintain an unmodified transmission matrix in order to propose a classical model with the closest resemblance to the ground truth hypothesis presented by the authors of the experiment.

	\section{\label{sec:validation}Validation tests}
	
	The results of the Jiuzhang experiments have been validated against a number of hypotheses and adversaries. These validations generally made use of three different tests: a Bayesian test, the Heavy Output Generation (HOG) test, and the comparison of the click cumulants of the different possible distributions with those of the experimental samples.
	
	In this section, we will compare the squashed states hypothesis and the squeezed ground truth hypothesis and we will determine which of them is better supported by the experimental data available (in the context of statistical hypothesis testing, this would tell us which hypothesis is the better explanation of the experiment). Moreover, we will use samples generated from the squashed states hypothesis to perform the HOG test~\footnote{All the data used in the computation of the validation tests is available upon reasonable request, or at \texttt{https://doi.org/10.5281/zenodo.7141021}.}.

	\subsection{\label{sec:cumulants}Click cumulants of the distribution}
	
	The comparison of click cumulants was first used in the validation of the Jiuzhang 2.0 experiment~\cite{zhong2021phase}. The authors used this method to investigate how robust the experimental samples are against classical simulation schemes based on marginal distributions. They claim that the presence of ``non-trivial genuine high-order correlation in the
	GBS samples are evidence of robustness against possible classical simulation schemes''~\cite{zhong2021phase}.  These correlation functions are given by cumulants~\cite{phillips2019benchmarking,fisher1932derivation,cardin2022photon} (also called Ursell~\cite{ursell_1927} functions, truncated correlation functions~\cite{duneau1973decrease} or cluster functions~\cite{duneau1973decrease}) defined in terms of moments of a multidimensional random variable $\bm{X} = (X_1,X_2,\ldots,X_M)$ as
	\begin{align}
	\begin{split}
		\kappa(X_1,\dots,X_n) =\sum_\pi& (|\pi|-1)!(-1)^{|\pi|-1}\\
		&\times\prod_{B\in\pi} \left\langle \prod_{i\in B}X_i \right\rangle,
	\end{split}
	\end{align}
	where $\pi$ runs through the list of all partitions of $\{ 1, ..., n \}$, $B$ runs through the list of all blocks of the partition $\pi$, and $|\pi|$ is the number of parts in the partition.
	Note that the first order cumulants are simply the means $\kappa(X_i) = \braket{X_i}$ and that the second order cumulants are the covariances $\kappa(X_i, X_j) = \braket{X_i X_j} - \braket{X_i} \braket{X_j}$.
	
	Since the probability distribution associated with a threshold detector experiment has binary outcomes $0,1$ in each mode, it is straightforward to see that moments of the distribution correspond to marginal probabilities
	\begin{align}
		\braket{X_{i_1}\ldots X_{i_n}} = \Pr\left(i_1=1,\ldots, i_n = 1 \right).
	\end{align}
	The latter in turn can be computed by constructing the marginal covariance matrix of the modes $i_1,\ldots,i_n$ and using Eq.~\eqref{eq:ground_truth_distribution}.

		\begin{figure*}[!htp]
		\centering
		\includegraphics[scale=0.36]{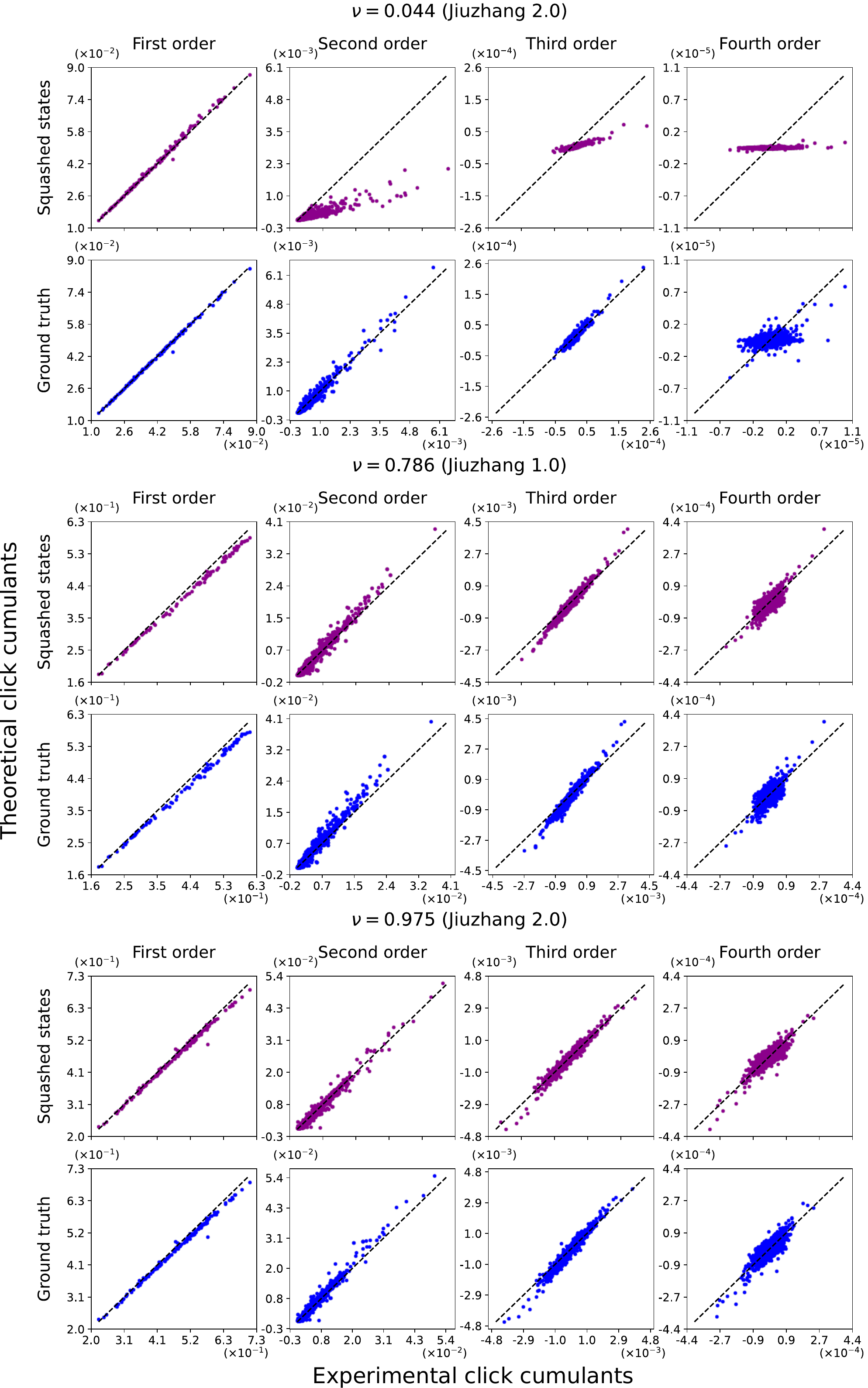}
		\caption{Comparison between the experimental click cumulants (up to fourth order) and those predicted by the squashed states (violet) and ground truth (blue) distributions for Jiuzhang 1.0 ($\nu=0.786$) and two configurations of Jiuzhang 2.0 ($\nu=0.044$ and $\nu=0.975$). For a hypothesis that perfectly describes the experimental samples, all the cumulants would lie in the dashed straight lines shown in the figure. The Pearson and Spearman correlation coefficients between theoretical and experimental cumulants are shown in Fig.~\ref{fig:coeff}.
		}
		\label{fig:click_cumulants}
	\end{figure*}

	\begin{figure*}[!ht]
		\centering
		\subfloat[]{\label{fig:pearson}
			\includegraphics[scale=0.39]{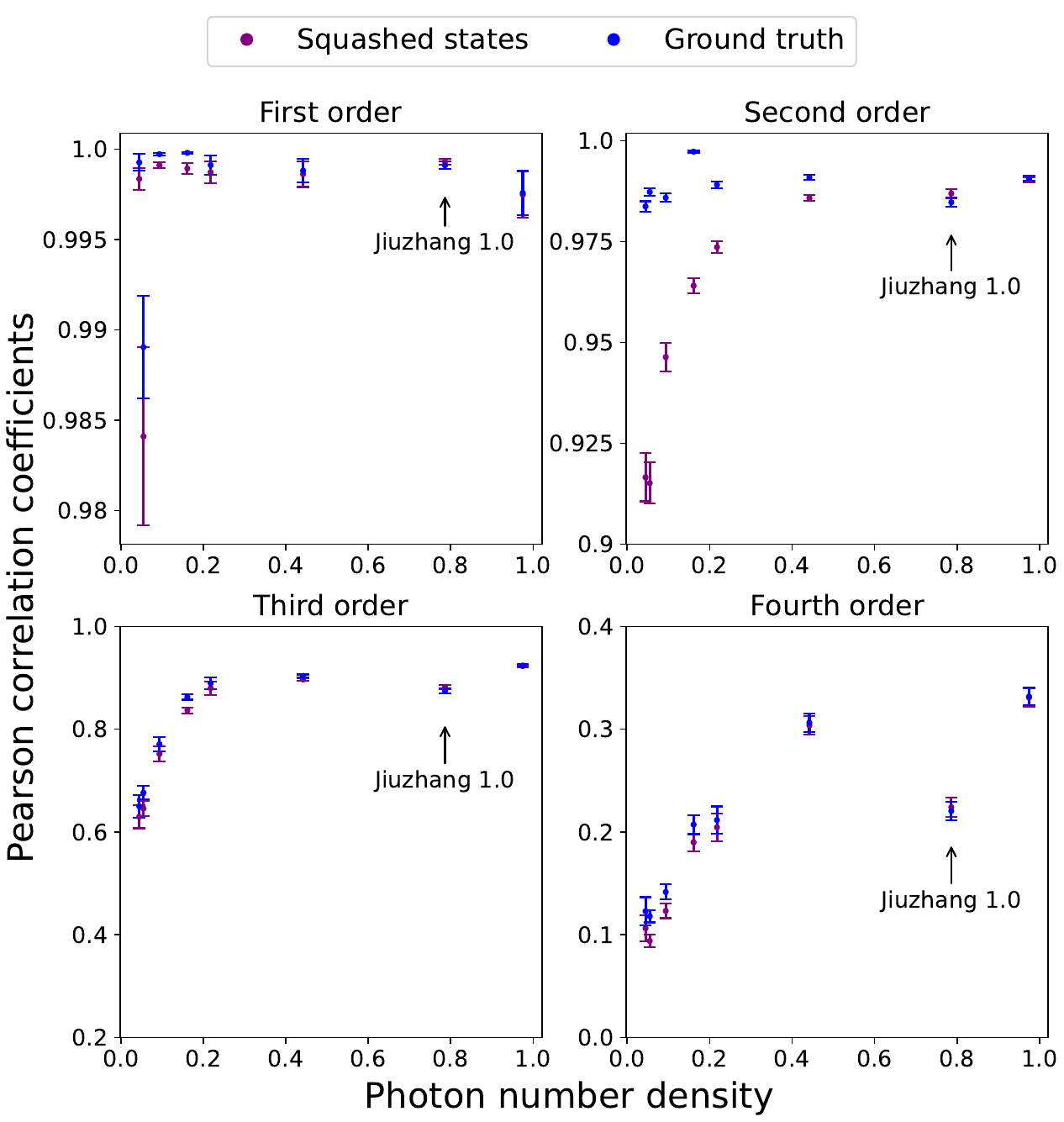}
		}
		\hfill
		\subfloat[]{\label{fig:spearman}
			\includegraphics[scale=0.39]{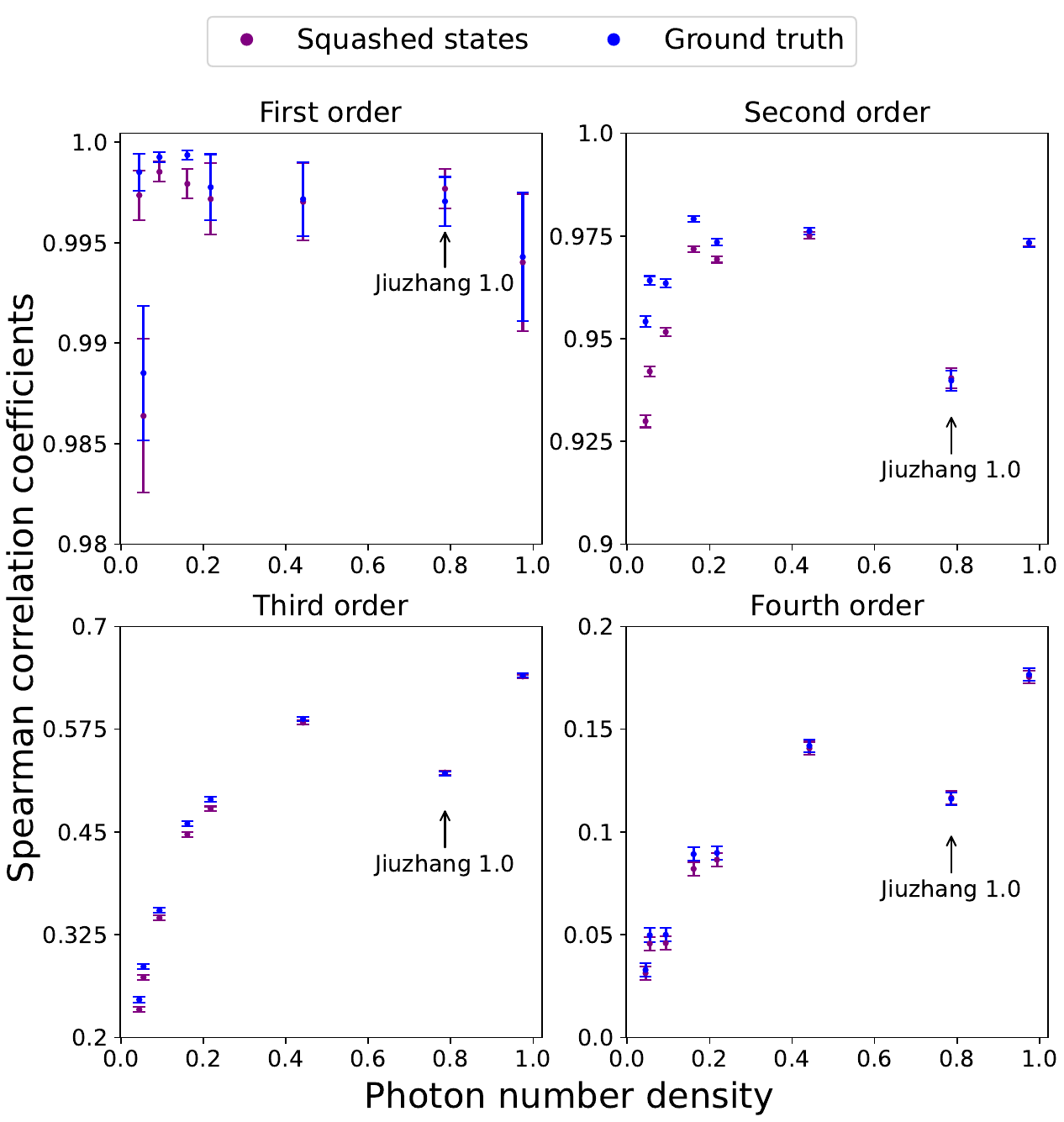}
		}
		\caption{Pearson (a) and Spearman (b) correlation coefficients between experimental click cumulants and those predicted by the theoretical squashed states (violet) and ground truth (blue) distributions as functions of the photon number density. The uncertainties in the correlation coefficients (represented by the error bars) were obtained using bootstrapping. Notice that, with increasing photon number density, the correlation coefficients obtained using the squashed states distribution become progressively closer to those obtained with the ground truth distribution.}
		\label{fig:coeff}
	\end{figure*}
	
	Following Ref.~\cite{zhong2021phase}, we compute the cumulants up to fourth order of the squashed states distribution of the different setups of the Jiuzhang experiment and compare them with those of the experimental samples.
	We also make the corresponding comparison with the ground truth distributions. For these computations, we used $10^7$ of the $\sim5\times10^7$ samples available in Ref.~\cite{ustc2020experimental}. Note that the number of cumulants increases sharply with the order; for a system with $M$ modes there are $M!/(\ell!(M-\ell)!)$ possible combinations of $\ell$ different modes (without repetitions), which is precisely the number of cumulants of order $\ell$. Considering this fact, we randomly select $10^5$ sets of three and four different modes for the computation of the third and fourth order cumulants, respectively. We compute all the possible cumulants of first and second order.
	
	Fig.~\ref{fig:click_cumulants} shows the results for Jiuzhang 1.0 ($\nu = 0.786$) and for two configurations of Jiuzhang 2.0 ($\nu = 0.044$ and $\nu = 0.975$). %
	Fig.~\ref{fig:coeff} shows the Pearson and Spearman correlation coefficients between experimental and theoretical cumulants as functions of $\nu$ (for all the configurations of Jiuzhang 1.0/2.0). The corresponding $p$-values, computed using the function \texttt{cor.test} from the R stats package~\cite{Rmanual}, are lower than \verb|machine_epsilon|$=2.2\times10^{-16}$ for every configuration, indicating a negligible probability of obtaining these correlation coefficients using uncorrelated data sets (i.e. a negligible probability of obtaining these correlation coefficients by chance). The same computation using the \texttt{pearsonr} and \texttt{spearmanr} functions from the SciPy stats (\texttt{scipy.stats}) module~\cite{2020SciPy-NMeth}, results in $p$-values that are effectively zero for all $\nu$.
	
	It is clear that for low densities, as exemplified by the $\nu = 0.044$ configuration in Fig.~\ref{fig:click_cumulants}, the cumulants predicted by the squashed states distribution are not consistent with the experimental results. In this case we can readily admit the ground truth of the experiment as the better hypothesis. This observation also holds for configurations with photon number densities between $\nu=0.055$ and $\nu=0.218$ as seen by looking at the Pearson and Spearman correlation coefficients in Fig.~\ref{fig:coeff}.
	However, we note that by increasing $\nu$, the cumulants of the squashed states distribution become progressively consistent with the experimental samples. Indeed, for setups with $\nu=0.786$ and $\nu=0.975$ (also for $\nu=0.442$), the experimental results are as consistent with the squashed states distribution as they are with the ground truth distribution of the experiment.
	In these cases, the comparison of cumulants does not allow to directly determine which of these two distributions better describes the experiment. This indicates that, for configurations with high photon number density, the presence of cumulants (up to fourth order) in the experimental samples can be efficiently reproduced by a classical hypothesis, which implies that the idea of having these correlations in the data does not necessarily shield the experiment from classical simulations. Moreover, this observation suggests that the comparison of click correlations between modes is not a completely reliable metric for the validation of GBS experiments.  
	
	This result is of particular importance even if it does not hold for all $\nu$. From Table~\ref{tab:experiment_specifics} (and also from Fig.~\ref{fig:click_probabilities} in Appendix~\ref{app:drummond}) it can be inferred that configurations with low $\nu$ have click number distributions that are mostly located below 40 clicks, that is, these configurations mostly generate samples with less than 40 clicks. This upper limit in the number of clicks makes the simulation of these setups a feasible task for a classical computer~\cite{quesada2018gaussian, kaposi2021polynomial, bulmer2022boundary}. It is the configurations of the Jiuzhang experiments with high photon number density that become increasingly difficult to simulate. The comparison of click cumulants shown here indicates that a simulation scheme using the squashed states hypothesis efficiently reproduces the results of the experiment in this regime. 
	
	Although, in principle, the comparison between theoretical and experimental cumulants could be carried out up to an arbitrary order, there is an important limitation that does not allow the direct comparison of correlations of fifth order or higher. As the authors of the experiment already noted (see Supplemental Material of Ref.~\cite{zhong2021phase}), experimental correlations of order equal or higher than five become increasingly difficult to distinguish from statistical noise. This is mostly due to computing cumulants with an insufficient number of experimental samples. However, the number of samples required to distinguish correlations from noise sharply increases with the order, which in turn, implies an increase of the time and memory needed for the computation of cumulants. 
	
	On account of this limitation, the authors of the experiment do not directly compare theoretical and experimental correlations of higher order. Instead, they use a statistical analysis to estimate the $p$-values of the Pearson correlation coefficients (between ground truth and experimental cumulants) up to 7th-order. Then, they extrapolate the estimated seed curve to compute $p$-values for higher orders. This allows the authors to claim that the $p$-values have statistical significance (i.e. $p<0.05$) up to order $(19\pm1)$, which implies the presence of 19th-order correlations in the experimental data (details of this statistical analysis can be found in the Supplemental Material of Ref.~\cite{zhong2021phase}). We are not convinced that this procedure allows to truly verify if theoretical and experimental correlations are compatible for higher orders. Consequently, we do not carry out the same statistical analysis using the squashed states distribution.
	
	We explain the compatibility between the squashed states and experimental cumulants for high photon-number density $\nu$ by noticing that the Jiuzhang experiments use input states with different squeezing parameters and threshold detectors. For a general GBS setup, it can be shown~\cite{grier2021complexity} that the expected mean number of clicks, $\bar{C}$, when using threshold detectors follows the relation  
	\begin{equation}
		\frac{1}{\bar{C}} = \frac{1}{\bar{N}} + \frac{1}{M},
		\label{eq:empirical_relation}    
	\end{equation}
	where $\bar{N}$ is the expected mean number of photons and $M$ is the number of output modes. In terms of the photon number density $\nu = \bar{N} / M$ we have
	\begin{equation}
		\frac{\bar{N}}{\bar{C}} = 1 + \nu.
		\label{eq:empirical_relation_2}    
	\end{equation}
	As can be seen from the equation above, $\bar{N}\sim \bar{C}$ for $\nu \ll 1$. This suggests that in this regime we can interpret each click detected as corresponding to approximately one single photon. On the other hand, for increasing $\nu$, $\bar{N} >\bar{C}$, implying that a single click corresponds to more than one photon (there is an increasing number of collisions). These observations suggest that in the low $\nu$ regime we can interpret a measurement using threshold detectors as an approximate photon number resolving (PNR) measurement. For increasing $\nu$ we can no longer state that the use of threshold detectors is similar to the use of PNR detectors. When the squeezing parameters of the input states are all the same, a PNR measurement can readily distinguish between squeezed and squashed states (this can be checked by comparing their second order cumulants, as has been done in a recent experiment by Madsen et al.~\cite{madsen2022quantum}). This does not necessarily hold when the squeezing parameters are different, $\nu$ is high or threshold detectors are used.

	\subsection{\label{sec:entropies}Bayesian test}

    \begin{figure*}[!htp]
		\centering
		\includegraphics[scale=0.41]{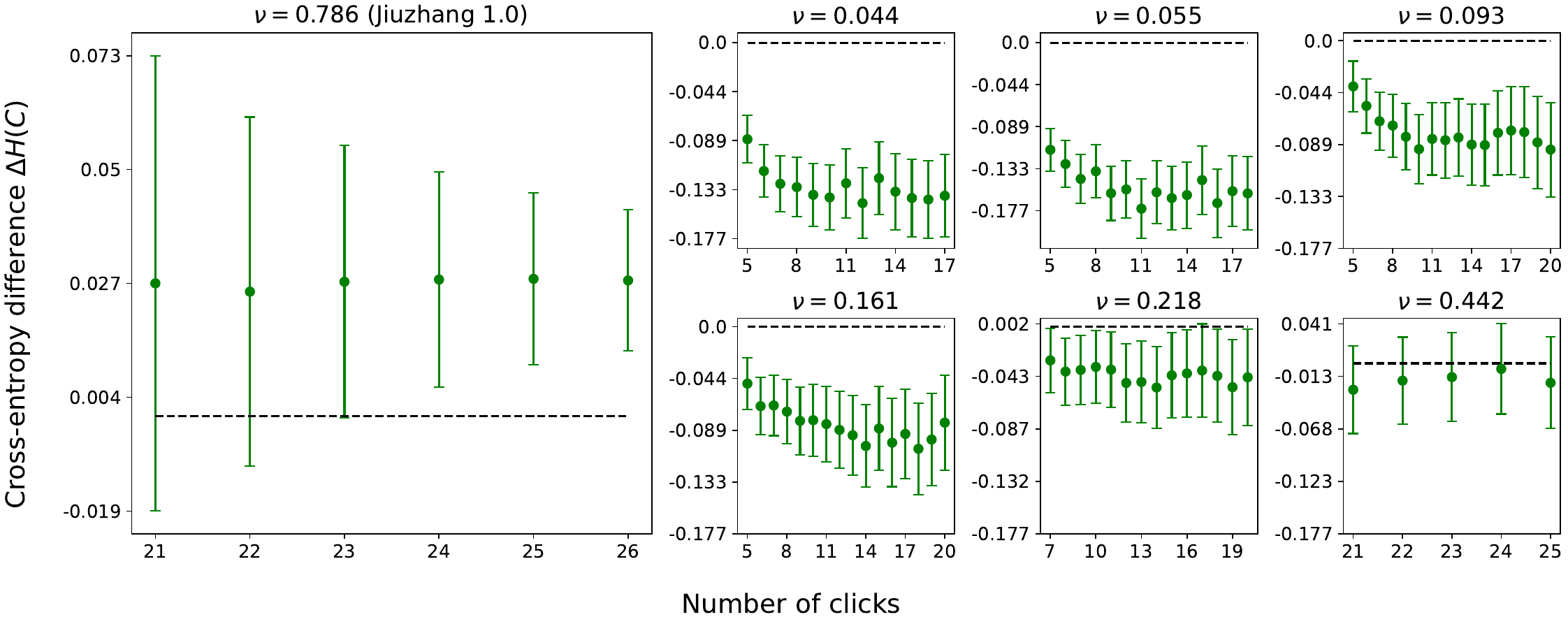}
		\caption{Bayesian test for the different setups of the Jiuzhang experiments in terms of $\Delta H(C)$. Error bars are obtained from the standard error of the mean and the computed uncertainty in the simulation of the grouped click probabilities. To compute $\Delta H(C)$ for configurations with $\nu\leq 0.442$, we used $4000$ experimental samples for each $C$. For the $\nu=0.786$ setup we used $L=\{4296, 7471, 12821, 21343, 34453, 53554\}$ samples for click numbers $C=\{21, 22, 23, 24, 25, 26\}$.}
		\label{fig:bay_entropies}
	\end{figure*}
	
	The Bayesian test that was used in the validation of the Jiuzhang 1.0/2.0 experiments compares two different hypotheses regarding the true probability distribution of the experimental samples~\cite{bentivegna2014bayesian}. This test gives the degree of confidence of one hypothesis over another. In this case, we compute the degree of confidence of the ground truth of the experiment over the squashed states hypothesis. 
	
	As was mentioned in Sec.~\ref{sec:introduction}, the Bayesian test relies on the computation of probabilities of individual samples, which is a computationally hard task for patterns with a high number of clicks. This means that the test cannot be used to validate the entirety of the experimental data and, consequently, does not represent a complete measure to readily verify if a GBS experiment achieves quantum computational advantage. Rather, the Bayesian test is used to
	build up confidence in the correct functioning of the GBS setup by ruling out possible classical hypotheses explaining the samples. 
	
	We will compute the test for samples with fixed click numbers (following Refs.~\cite{zhong2020quantum, zhong2021phase, villalonga2021efficient}) between 5 and 26. This upper limit in the number of clicks takes into account that the computation of the corresponding probabilities requires quadruple precision of complex type numbers, thus increasing the computational cost. Indeed, obtaining the probability for a single pattern of 25 clicks takes $\sim3.5$ hours using a 64-core CPU with two AMD Rome 7532 processors with $2.4\,\text{GHz}$ clock speed, using a custom implementation of the Torontonian function (see Ref.~\cite{polyq2022tor}) which uses Quadruple-precision (128 bits for each real number) provided by the library \texttt{DoubleFloats.jl}~\cite{sarnoffDoubleFloats2022} in the Julia Programming language~\cite{bezanson2017julia}.
	
	Consider a set $\bm{S} = \{\bm{s}_1,\dots,\bm{s}_L\}$ of $L$ experimental samples, each of them containing $C$ clicks. The probability of obtaining one of these samples, given that it has $C$ clicks, under the hypothesis $\text{HYP} \in \{\text{SQUA}, \text{\text{SQUE}}\}$ is given by
	\begin{align}
		\mathrm{Pr}_{(\text{HYP})}(\bm{s}_k|C) =  \mathrm{Pr}_{(\text{HYP})}(\bm{s}_k) / \mathrm{Pr}_{(\text{HYP})}(C)    
	\end{align}
	where $\mathrm{Pr}_{(\text{HYP})}(\bm{s}_k)$ is the probability of sample $\bm{s}_k$ under hypothesis $\text{HYP}$ given in terms of a Torontonian (cf. Eq.~\eqref{eq:ground_truth_distribution}) and $\mathrm{Pr}_{(\text{HYP})}(C)$ is the grouped probability of obtaining $C$ clicks in total, again under the hypothesis $\text{HYP}$. The probability of obtaining the set of samples $\bm{S}$ under a given hypothesis $\text{HYP}$ takes the form
	\begin{equation}
		\mathrm{Pr}_{(\text{HYP})}(\bm{S}|C) = \prod_{k=1}^L \mathrm{Pr}_{(\text{HYP})}(\bm{s}_k|C).
		\label{eq:joint_gt_probability}
	\end{equation}
	
	We define the Bayesian ratio, $r_\text{B}(C)$, which can be interpreted as the probability assigned to the ground truth hypothesis for a given number of clicks, as
	\begin{align}
		\begin{split}
			r_\text{B}(C)&=\frac{\mathrm{Pr}_{(\text{SQUE})}(\bm{S}|C)}{\mathrm{Pr}_{(\text{SQUE})}(\bm{S}|C) + \mathrm{Pr}_{(\text{SQUA})}(\bm{S}|C)} \\
			&= \frac{1}{1 + \chi_\text{B}(C)},
		\end{split}
		\label{eq:bayesian_ratio}    
	\end{align}
	where $\chi_\text{B}(C) = \mathrm{Pr}_{(\text{SQUA})}(\bm{S}|C)/ \mathrm{Pr}_{(\text{SQUE})}(\bm{S}|C)$. The Bayesian test consists in checking the convergence of $r_\text{B}(C)$ when the number of samples is increased: if $r_B(C) \rightarrow 1$ for any $C$, we conclude that the ground truth hypothesis is more likely to describe the experimental samples. Conversely, if $r_B(C) \rightarrow 0$ for any $C$, the squashed states hypothesis becomes more likely.

	An alternative way to express this test is obtained by writing $\chi_\text{B}= \exp(L\Delta H (C))$, where
	\begin{align}
		\begin{split}
			\Delta H(C) =& -\frac{1}{L}\sum_{k=1}^L\ln\left[\mathrm{Pr}_{(\text{SQUE})}(\bm{s}_k|C)\right]\\
			&+\frac{1}{L}\sum_{k=1}^L\ln\left[\mathrm{Pr}_{(\text{SQUA})}(\bm{s}_k|C)\right]\\
			=& H_{(\text{SQUE})}(C) - H_{(\text{SQUA})}(C).
		\end{split}
		\label{eq:bay_cross_ent_diff}
	\end{align}
	The quantities $H_{(\text{SQUE})}$ and $H_{(\text{SQUA})}$ are estimators of the cross-entropy, for a given number of counts, of the ground truth and squashed states distributions relative to the real probability distribution of the experimental samples. In terms of the cross-entropy difference $\Delta H(C)$, $r_\text{B}(C) = \left[1+\exp(L\Delta H(C))\right]^{-1}$ and, for a increasing number of samples, the condition $r_B(C) \rightarrow 1$ is equivalent to $\Delta H(C)<0$, while $r_B(C) \rightarrow 0$ is equivalent to $\Delta H(C)>0$.

	An important step for the computation of the Bayesian test is the determination of the grouped click probability distributions $\mathrm{Pr}_{(\text{SQUE})}(C)$ and $\mathrm{Pr}_{(\text{SQUA})}(C)$. To do this, we use the simulation method introduced in Ref.~\cite{drummond2022simulating}. The definition of this method, as well as the resulting click probability distributions and the parameters used in the simulation, are shown in Appendix~\ref{app:drummond}. It is worth mentioning that this method allows the determination of $\mathrm{Pr}_{(\text{SQUE})}(C)$ and $\mathrm{Pr}_{(\text{SQUA})}(C)$ for all click numbers $C$, not only those for which the probabilities of individual samples are easily computed.

	Fig.~\ref{fig:bay_entropies} shows the results of the Bayesian test for the Jiuzhang 1.0 and Jiuzhang 2.0 experiments in terms of cross-entropy differences. For configurations with $\nu\leq0.218$, we considered experimental samples with click numbers between 5 and 20. In these cases, we used $4000$ samples for each $C$. For the $\nu = 0.786$ and $\nu = 0.442$ setups, we considered click numbers between $21$ and $26$ due to the lack of sufficient samples (less than $1000$ for each click number) with $C$ lower than $20$. In the case of $\nu = 0.786$, we used $L=\{4296, 7471, 12821, 21343, 34453, 53554\}$ samples for click numbers $C=\{21, 22, 23, 24, 25, 26\}$. For $\nu = 0.442$, we used $4000$ samples for each $C$. The results for $\nu=0.975$ are not addressed on account of the insufficient number of experimental samples within the range of click numbers considered. Indeed, for this configuration, there were more than $L=1000$ samples only for click numbers starting from $C=30$. Computing probabilities for samples with this number of clicks was beyond our computational capabilities. 
	
	For the different configurations of the Jiuzhang 2.0 experiment, including the second brightest configuration ($\nu=0.442$), the Bayesian test consistently indicates that the ground truth is more likely to explain the experimental samples than the squashed states hypothesis. However, as was the case with the click cumulants comparison, by increasing $\nu$, the values of $\Delta H(C)$ in the range of the click numbers considered are lower than zero by a progressively smaller margin, indicating that the degree of confidence in the squashed states hypothesis increases. For the $\nu=0.786$ configuration of Jiuzhang 1.0, the Bayesian test indicates that the squashed states hypothesis is more likely than the ground truth of the experiment. 
	
	The discrepancy between the results for the configurations $\nu=0.442$ and $\nu=0.786$, both belonging to the high photon number density regime, may be explained by the refinement of the light source of the Jiuzhang 2.0 experiment with respect to Jiuzhang 1.0. Indeed, a better preparation of the input two-mode squeezed states directly challenges the idea that the input states of the interferometers are better described by squashed states, which, as was mentioned earlier, are approximations to squeezed states with losses in their preparation.

	\subsection{\label{sec:hog}HOG test}

    \begin{figure*}[!htp]
		\centering
		\includegraphics[scale=0.41]{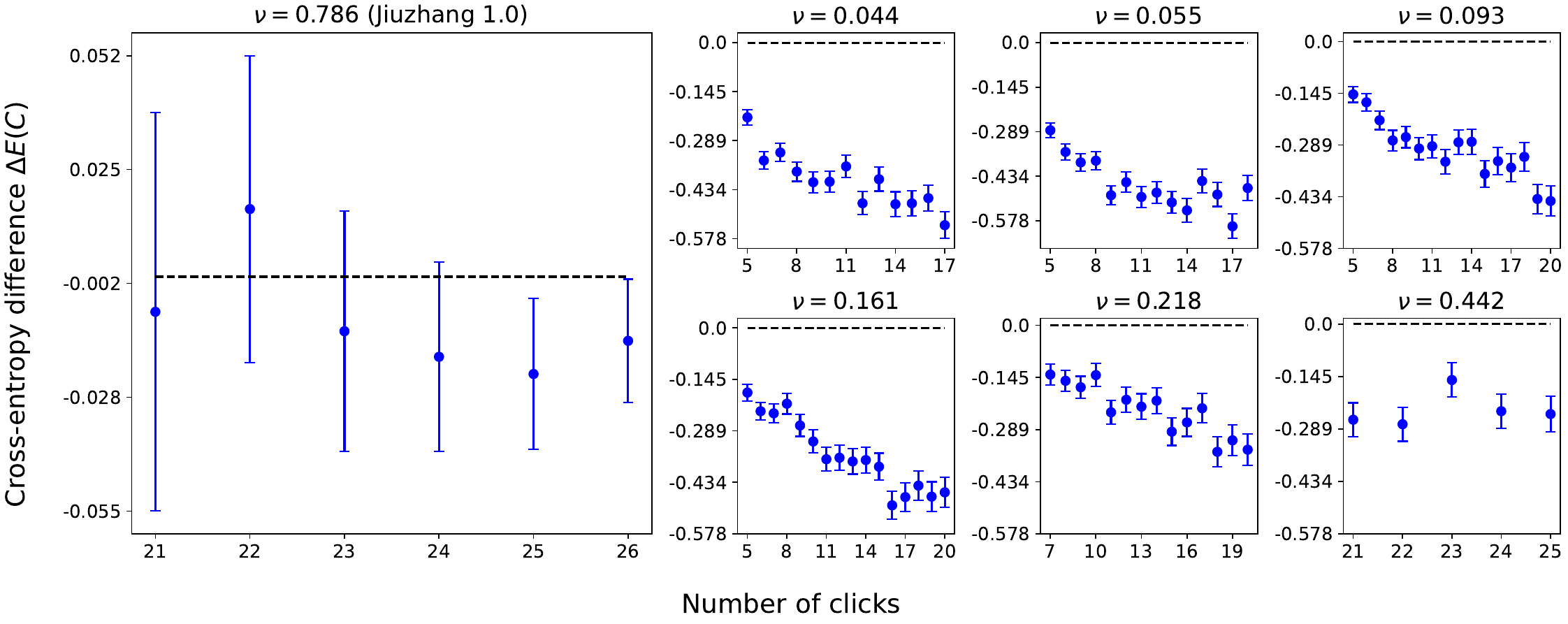}
		\caption{HOG test for the different setups of the Jiuzhang experiments in terms of $\Delta E(C)$. Error bars are obtained from the standard error of the mean and the computed uncertainty in the simulation of the grouped click probabilities. To compute $\Delta E(C)$ for configurations with $\nu\leq 0.442$, we used $4000$ experimental and squashed states samples for each $C$. For the $\nu=0.786$ setup we used $L=\{4296, 7471, 12821, 21343, 34453, 53554\}$ samples for click numbers $C=\{21, 22, 23, 24, 25, 26\}$.}
		\label{fig:hog_entropies}
	\end{figure*}

	The Heavy Output Generation (HOG) test was first introduced in Ref.~\cite{zhong2020quantum} for the validation of the Jiuzhang 1.0 experiment. This test compares the probabilities with respect to the ground truth distribution of two sets of samples: one corresponding to experimental samples, and the other corresponding to samples from an alternative distribution, which in our case corresponds to the squashed states distribution. This test verifies if a sampler using the alternative distribution can generate samples with higher ground truth probability than the experimental samples.
	
    By relying on the computation of probabilities of individual samples, the HOG test has the same computational limitations as the Bayesian test. Namely, it can only be calculated for patterns with a small number of clicks. This means that the HOG test cannot be used to completely verify a GBS quantum advantage claim; it can only be used to rule out possible classical samplers and increase the confidence in the correct functioning of the experiment.

	The definition of the HOG test is made in a similar fashion to that of the Bayesian test. We define the HOG ratio, $r_\text{HOG}(C)$, as
	\begin{align}
		\begin{split}
			r_\text{HOG}(C)&=\frac{\mathrm{Pr}_{(\text{SQUE})}(\bm{S}|C)}{\mathrm{Pr}_{(\text{SQUE})}(\bm{S}|C) + \mathrm{Pr}_{(\text{SQUE})}(\bm{S'}|C)} \\
			&= \frac{1}{1 + \chi_\text{HOG}(C)},
		\end{split}
		\label{eq:hog_ratio}    
	\end{align}
	where $\chi_\text{HOG}(C) = \mathrm{Pr}_{(\text{SQUE})}(\bm{S'}|C)/ \mathrm{Pr}_{(\text{SQUE})}(\bm{S}|C)$ and $\bm{S'}=\{\bm{s'}_1,\dots,\bm{s'}_L\}$ is a set of $L$ samples obtained from the squashed states distribution (each one with $C$ clicks). $\mathrm{Pr}_{(\text{SQUE})}(\bm{S}|C)$ is computed according to Eq.~\eqref{eq:joint_gt_probability}. We emphasize that in the last equation all the probabilities are taken with respect to the squeezing distribution (SQUE).  The test consists in checking the convergence of $r_\text{HOG}(C)$ for an increasing number of samples: $r_\text{HOG}(C)\rightarrow 1$ if we generally find that the experimental samples have higher ground truth probability for any $C$, while $r_\text{HOG}(C)\rightarrow 0$ when the samples from the squashed states distribution have higher ground truth probability for any $C$.
	
	As in the case of the Bayesian test, the ratio $\chi_\text{HOG}(C)$ may be rewritten as $\chi_\text{HOG}(C) = \exp(L\Delta E(C))$, where
	\begin{align}
		\begin{split}
			\Delta E(C) =& -\frac{1}{L}\sum_{k=1}^L\ln\left[\mathrm{Pr}_{(\text{SQUE})}(\bm{s}_k|C)\right]\\
			&+\frac{1}{L}\sum_{k=1}^L\ln\left[\mathrm{Pr}_{(\text{SQUE})}(\bm{s'}_k|C)\right]\\
			=& E_{(\text{SQUE})}(C) - E'_{(\text{SQUE})}(C).
		\end{split}
		\label{eq:hog_cross_ent_diff}  \end{align}

	In terms of the cross-entropy difference $\Delta E(C)$~\cite{villalonga2021efficient}, $r_\text{HOG}(C) = \left[1 + \exp(L\Delta E(C))\right]^{-1}$ and, for increasing $L$, the conditions  $r_\text{HOG}(C)\rightarrow 1$ and $r_\text{HOG}(C)\rightarrow 0$ are equivalent to $\Delta E(C)<0$ and $\Delta E(C)>0$, respectively.

	In Fig.~\ref{fig:hog_entropies} we show the results for the HOG test, in terms of $\Delta E(C)$, for the Jiuzhang 1.0 and Jiuzhang 2.0 experiments. The parameters used in the calculation of the grouped click probabilities were the same as those used in the computation of the Bayesian test (see Appendix~\ref{app:drummond}). The ranges of click numbers considered here, as well as the corresponding $L$, were also the same as in the case of the Bayesian test. The results for the $\nu =0.975$ configuration of Jiuzhang 2.0 are not shown due to the lack of experimental samples in the range of click numbers considered.

	As can be seen in Fig.~\ref{fig:hog_entropies}, for the Jiuzhang 1.0 experiment, we get inconclusive results, as $\Delta E(C)>0$ for $C=22$ but has a negative sign for the remaining click numbers (for the test to be conclusive $\Delta E(C)$ must have the same sign for all the $C$ considered). For all the configurations of Jiuzhang 2.0, we consistently find that the experimental samples have higher ground truth probability than the squashed states samples, even for setups in the high photon number density regime, in agreement with the results of the Bayesian test. Additionally, contrary to the case of the Bayesian test, the cross-entropy difference $\Delta E(C)$ does not differ from zero by a progressively smaller margin with increasing $\nu$.

    It is important to mention that although the present model cannot generate samples that outperform the experimental results at the HOG test, a modified version of the squashed states hypothesis, which includes corrections to the input mean photon numbers and the transmission matrix of the interferometer (as proposed in Ref.~\cite{drummond2022simulating}), might be able to generate classical samples with higher ground truth probability than the experimental samples. Additionally, this modified model might also lead to more accurate results for the Bayesian test and the comparison of click cumulants.

	\section{\label{sec:discussion}Discussion}
	
	In this work we proposed an alternative hypothesis for the validation of the Jiuzhang GBS experiments. This hypothesis is based on the probability distribution of classical mixtures of coherent states that we call squashed states.
	For the validation of the experimental results against this alternative hypothesis, we used the same methods as the authors of the experiments.
	
	The results for the click cumulants comparison show that the theoretical cumulants predicted by the squashed states distribution are not consistent with the experimental results for setups with low photon number density, $\nu$. However, it is noticeable that with increasing $\nu$ the squashed states cumulants become progressively consistent with those obtained from the experimental samples. This trend is confirmed by the results for the configurations with high photon number density, for which the theoretical cumulants predicted by the squashed states distribution are as compatible with the experimental results as those predicted by the ground truth of the experiment. These results suggest that the presence of cumulants (up to fourth order) in the experimental samples can be efficiently reproduced by classical hypotheses lacking any quantum correlation.
	
	The results of the Bayesian test show differences with respect to those of the click cumulants comparison. By analyzing the cross-entropy difference $\Delta H(C)$ as a function of the number of clicks, we consistently find that, for the different configurations of the Jiuzhang 2.0 experiment, $\Delta H(C)<0$, indicating that the ground truth of the experiment is more likely to describe the experimental samples than the squashed states hypothesis. However, we note the tendency of $\Delta H(C)$ to approach zero with increasing photon number density, suggesting an increasing degree of confidence in squashed states distribution with respect to the ground truth. On the other hand, for the Jiuzhang 1.0 experiment, we find that $\Delta H(C)>0$, which indicates that the squashed states hypothesis is a better explanation of the experimental samples than the ground truth of the experiment. The contrast between the results of the Bayesian test for the configurations with $\nu=0.442$  and $\nu=0.786$, both being configurations with high photon number density, may be explained by the improvement of the light source of the Jiuzhang 2.0 relative to Jiuzhang 1.0.

	We generated samples from the squashed states distribution and compared them against the experimental samples at the task of heavy output generation (HOG) from the squeezed states distribution. In this case we found that, except for a subset of the data in Jiuzhang 1.0, the experiments perform better at the HOG task than the samples from the squashed states hypothesis.
	
	Our results provide a more nuanced picture of the different tools employed in arguing about quantum advantage in threshold GBS experiments. On one hand, we found that the presence of correlations up to fourth order in the experimental samples from Jiuzhang 1.0 and 2.0 in the high photon number density regime can be efficiently reproduced by states lacking any quantum features.
	On the other hand, we observe that, for some of the setups in this high $\nu$ regime, the squashed states hypothesis is not the better hypothesis for the explanation of the experimental samples.
	This result indicates that the comparison of correlation functions is not entirely reliable for the validation of quantum advantage claims in threshold GBS experiments. 
	
	Moreover, the samples generated using the squashed states hypothesis cannot spoof the HOG test when compared to the samples of the experiment, even for the configuration for which the Bayesian test indicates that the squashed states hypothesis is more likely to explain the experimental data (namely, for the Jiuzhang 1.0 experiment). Note, however, that other efficient classical methods have already outperformed the Jiuzhang 1.0 and 2.0 at this task~\cite{villalonga2021efficient}. Nevertheless, in absence of rigorous results about the significance of the HOG test for GBS (as opposed to RCS), the inconclusive result for Jiuzhang 1.0 must be taken with a grain of salt, especially when alternative hypotheses are more probable than the ground truth.
	
    Our work provides a new adversary that should be considered against future GBS experiments and, perhaps more importantly, further motivates the need to identify proper metrics and optimal classical adversaries for quantum advantage in the context of threshold GBS.

	\section*{Acknowledgements}
	We wish to thank  M.-C. Chen, Y.-H. Deng, Y.-C. Gu, H. Su, J. Qin, H.-S. Zhong, and C.-Y. Lu for help in the validation of some of the results in this preprint and pointing out an error in the calculation of the Bayesian test for the modified squashed state hypothesis discussed in previous version of this manuscript (v1 and v2 on the arXiv). N.Q. acknowledges support from the Ministère de l’Économie et de l’Innovation du Québec and the Natural Sciences and Engineering Research Council of Canada. This research was enabled in part by support provided by the Digital Research Alliance of Canada (formerly Compute Canada). N.Q. and J.M.-C. thank B. Villalonga and J.F.F. Bulmer for helpful pointers in parsing the data in Refs. ~\cite{ustc2020experimental,ustc2021raw}, P. Drummond for insightful correspondence and M. Houde for a critical reading of the manuscript.

	\appendix
 
	\section{\label{app:tor}The Torontonian}
	As discussed in Sec.~\ref{sec:distributions}, the ground truth and squashed states distributions depend on the Torontonian~\cite{quesada2018gaussian} of a matrix constructed from the corresponding covariance matrices of the output states of the interferometer. For a $2N \times 2N$ matrix $\bm{A}$, the Torontonian, $\text{Tor}(\bm{A})$, is computed as
	\begin{equation}
		\text{Tor}(\bm{A}) = (-1)^N \sum_{Z \in \Omega([N])}\frac{(-1)^{|Z|}}{\sqrt{\text{det}\left[\left(\mathbb{I}-A\right)_{(Z)}\right]}}.
		\label{eq:torontonian}    
	\end{equation}
	Here, $\Omega([N])$ is the power set, i.e., the set of all subsets, of $[N] = \{1,2,\dots,N\}$; $|Z|$ stands for the cardinality of set $Z$, that is, the number of elements in $Z$; and $\mathbb{I}$ is the $2M\times2M$ identity matrix. If we write $Z = \{z_1,\dots,z_n\}$, with $n=|Z|$, then $\bm{A}_{(Z)}$ is the matrix obtained from removing the rows and columns $\{z_1,\dots,z_n\}$ and $\{z_1 + N,\dots,z_n+N\}$ from matrix $\bm{A}$.

	\section{\label{app:drummond}Simulation of grouped click probabilities}

	In this appendix we describe the positive $P$-distribution simulation method~\cite{drummond2022simulating} used in Secs.~\ref{sec:entropies} and \ref{sec:hog} for the computation of the ground truth and squashed states grouped click probability distributions.
	
	Suppose that the input state of the interferometer in the Jiuzhang experiments is represented by the density operator $\rho_\text{in}$. This state can be expanded in terms of its positive $P$-representation~\cite{drummond2014quantum}, $P(\bm{\alpha},\bm{\beta})$, which is a phase space representation over a subspace of the complex plane. The expansion reads
	\begin{equation}
		\rho_\text{in} =  \text{Re}\left[\int P(\bm{\alpha},\bm{\beta})\frac{|\bm{\alpha}\rangle\langle\bm{\beta^*}|}{\langle\bm{\alpha}|\bm{\beta^*}\rangle}d\mu(\bm{\alpha},\bm{\beta})\right],
		\label{eq:positive_p_rep_input}    
	\end{equation}
	where $|\bm{\alpha}\rangle=|\alpha_1,\dots,\alpha_{K}\rangle$ and $|\bm{\beta}\rangle=|\beta_1,\dots,\beta_{K}\rangle$ are $K$-mode coherent states, with $K$ being the number of input states of the interferometer.  $d\mu(\bm{\alpha},\bm{\beta})$ corresponds to an integration measure over the $2K$-dimensional complex space of the amplitudes $\bm{\alpha}=(\alpha_1,\dots,\alpha_{K})$ and $\bm{\beta}=(\beta_1,\dots,\beta_{K})$.
	
	The state of the system after the interferometer, $\rho_\text{out}$, has a similar representation in terms of the same $P$-distribution, we need only modify the multimode coherent states in Eq.~\eqref{eq:positive_p_rep_input}:
	\begin{equation}
		\rho_\text{out} =  \text{Re}\left[\int P(\bm{\alpha},\bm{\beta})\frac{|\bm{\bar{\alpha}}\rangle\langle\bm{\bar{\beta}^*}|}{\langle\bm{\bar{\alpha}}|\bm{\bar{\beta}^*}\rangle}d\mu(\bm{\alpha},\bm{\beta})\right].
		\label{eq:positive_p_rep_output}    
	\end{equation}
	Here, $\bm{\bar{\alpha}}=\bm{T}\bm{\alpha}$ and $\bm{\bar{\beta}}=\bm{T^*}\bm{\beta}$,  where $\bm{T}$ is the $M\times K$ transfer matrix representing the action of the interferometer, where $M$ is the number of output modes. Notice that $|\bm{\bar{\alpha}}\rangle$ and $|\bm{\bar{\beta}}\rangle$ become now $M$-mode coherent states.
	
	With this representation of the output state, we can compute the probability of obtaining a detection pattern $\bm{s}=(s_1,\dots,s_M)$, with $s_i=0,1$, (when the system is measured using threshold detectors) as
	\begin{align}
			&\mathrm{Pr}(\bm{s}) = \text{Tr}\left(\rho_\text{out}\Pi_{\bm{s}}\right)\\
            &\quad =\text{Re}\Biggl[\int P(\bm{\alpha},\bm{\beta})
            \textrm{Tr}\left(\frac{|\bm{\bar{\alpha}}\rangle\langle\bm{\bar{\beta}^*}|}{\langle\bm{\bar{\alpha}}|\bm{\bar{\beta}^*}\rangle}\Pi_{\bm{s}}\right)d\mu(\bm{\alpha},\bm{\beta})\Biggr]. \nonumber
		\label{eq:det_probability}
	\end{align}
	$\Pi_{\bm{s}}$ is the operator representing the measurement of pattern $\bm{s}$, which can be expressed as
	\begin{equation}
		\Pi_{\bm{s}} = \bigotimes_{k\in\mathcal{V}(\bm{s})}|0_k\rangle\langle0_k|\bigotimes_{j\in\mathcal{C}(\bm{s})}\left(\mathbb{I}^{(j)}-|0_j\rangle\langle0_j|\right),
		\label{eq:det_operator}    
	\end{equation}
	where $|0_k\rangle$ is the vacuum state in the Hilbert state of mode $k$ and $\mathbb{I}^{(k)}$ the corresponding identity operator. The set $\mathcal{C}(\boldsymbol{s})$ is composed by the modes where a click is detected when the result of the measurement is pattern $\bm{s}$. Conversely, the elements of set $\mathcal{V}(\boldsymbol{s})$ are the modes where no click was detected. After a direct computation of the trace, it can be shown that
	\begin{equation}
		\mathrm{Pr}(\bm{s}) = \text{Re}\left[\int P(\bm{\alpha},\bm{\beta})\Lambda(\bm{s},\bm{\bar{\alpha}},\bm{\bar{\beta}})\,d\mu(\bm{\alpha},\bm{\beta})\right],
		\label{eq:det_probability_2}    
	\end{equation}
	where
	\begin{equation}
		\Lambda(\bm{s},\bm{\bar{\alpha}},\bm{\bar{\beta}})=\prod_{k=1}^Me^{-\bar{\alpha}_k\bar{\beta}_k}\left(e^{\bar{\alpha}_k\bar{\beta}_k}-1\right)^{s_k}.
		\label{eq:lambda_term}    
	\end{equation}

	Now, suppose that $\mathcal{D}(C)$ is the set of all detection patterns $\bm{s}$ with $C$ clicks. Then, the probability of obtaining $C$ clicks, $\mathrm{Pr}(C)=\sum_{\bm{s}\in\mathcal{D}(C)}\mathrm{Pr}(\boldsymbol{s})$, takes the form
	
	\begin{align}
	\begin{split}
		\mathrm{Pr}(C) =\text{Re}&\Biggl[\int P(\bm{\alpha},\bm{\beta})\\
		&\times\sum_{\bm{s}\in\mathcal{D}(C)}\Lambda(\bm{s},\bm{\bar{\alpha}},\bm{\bar{\beta}})\,\,d\mu(\bm{\alpha},\bm{\beta})\Biggr].
		\label{eq:click_probability}    
	\end{split}
	\end{align}
	
	By defining the function
	\begin{equation}
		G(\bm{\bar{\alpha}},\bm{\bar{\beta}};\theta_l) = \prod_{j=1}^M\left[e^ {-\bar{\alpha}_j\bar{\beta}_j} + e^{i\theta_l}\left(1-e^ {-\bar{\alpha}_j\bar{\beta}_j}\right)\right],
		\label{eq:fourier_coefficient}    
	\end{equation}
	where $\theta_l=2\pi l/(M+1)$, we can write
	\begin{align}
		\begin{split}
			\sum_{\bm{s}\in\mathcal{D}(C)}\Lambda(\bm{s},\bm{\bar{\alpha}},\bm{\bar{\beta}}) &= \frac{1}{M+1}\sum_{l=1}^{M+1}G(\bm{\bar{\alpha}},\bm{\bar{\beta}};\theta_l)e^{-iC\theta_l} \\
			&= \tilde{G}(\bm{\bar{\alpha}},\bm{\bar{\beta}};C),
			\label{eq:fourier_transform}
		\end{split}
	\end{align}
	and thus, the probability of obtaining $C$ clicks can be simply written as
	\begin{equation}
		\mathrm{Pr}(C) =\text{Re}\left[\int P(\bm{\alpha},\bm{\beta})\tilde{G}(\bm{\bar{\alpha}},\bm{\bar{\beta}};C)d\mu(\bm{\alpha},\bm{\beta})\right].
		\label{eq:click_probability_fourier}
	\end{equation}
	The previous expression does not depend on the number elements of $\mathcal{D}(C)$, which is generally large for a large number of modes. This fact makes the computation of $\mathrm{Pr}(C)$ a feasible task for a classical computer.

	Notice that obtaining $\mathrm{Pr}(C)$ is equivalent to the computation of a sort of expected value of the function $\tilde{G}(\bm{\bar{\alpha}},\bm{\bar{\beta}};C)$ with respect to distribution $P(\bm{\alpha},\bm{\beta})$. Thus, we can estimate $\mathrm{Pr}(C)$ numerically by generating $\mathcal{N}$ pairs of random vectors $\bm{\alpha_k}=(\alpha_1^{(k)},\dots,\alpha_K^{(k)})$ and $\bm{\beta}_k=(\beta_1^{(k)},\dots,\beta_K^{(k)})$ distributed as $P(\bm{\alpha},\bm{\beta})$; then, computing the corresponding $\tilde{G}(\bm{\bar{\alpha}}_k,\bm{\bar{\beta}}_k;C)$ functions; and finally, taking the real part of the average $\frac{1}{\mathcal{N}}\sum_{k=1}^{\mathcal{N}}\tilde{G}(\bm{\bar{\alpha}}_k,\bm{\bar{\beta}}_k;C)$. The random variables $\bm{\alpha}_k$ and $\bm{\beta}_k$ will depend only on the initial state in consideration.
	
	For an input state of $K$ single-mode squeezed states (SMSS), we can generate the samples $\bm{\alpha}_k$ and $\bm{\beta}_k$ in the following way~\cite{drummond2022simulating}. First, we calculate the mean photon numbers $\bar{n}_j = \langle a_j^\dagger a_j\rangle=\sinh^2(r_j)$ and coherences $\bar{m}_j = \langle a_j^2\rangle = \frac{1}{2}\sinh(2r_j)$ of the $K$ input states. Here $a_j^\dagger$ and $a_j$ stand for the creation and annihilation operators of mode $j$ and $r_j$ is the squeezing parameter corresponding to that same mode. We then generate $\mathcal{N}$ pairs of real Gaussian random vectors $\bm{u}_k=(u_1^{(k)},\dots,u_K^{(k)})$ and $\bm{v}_k=(v_1^{(k)},\dots,v_K^{(k)})$. The components of these vectors must satisfy the relations $\mathbb{E}(u_j^{(k)}v_j^{(k)}) = 0$, $\mathbb{E}(u_j^{(k)}u_l^{(k)}) = \mathbb{E}(v_j^{(k)}v_l^{(k)})=\delta_{jl}$, where $\mathbb{E}(\cdot)$ indicates the expectation value with respect to the Gaussian distribution. Finally, with this definition of $\bm{u}_k$ and $\bm{v}_k$, we can write the components of $\bm{\alpha}_k$ and $\bm{\beta}_k$ as
	\begin{align}
		\begin{split}
			\alpha_j^{(k)} = u_j^{(k)}\sqrt{\tfrac{1}{2}(\bar{n}_j+\bar{m}_j)} + iv_j^{(k)}\sqrt{\tfrac{1}{2}(\bar{n}_j-\bar{m}_j)},\\
			\beta_j^{(k)} = u_j^{(k)}\sqrt{\tfrac{1}{2}(\bar{n}_j+\bar{m}_j)} - iv_j^{(k)}\sqrt{\tfrac{1}{2}(\bar{n}_j-\bar{m}_j)}.
		\end{split}
		\label{eq:sampling_amplitudes}
	\end{align}

	\begin{figure*}[ht!]
		\centering
		\includegraphics[scale=0.50]{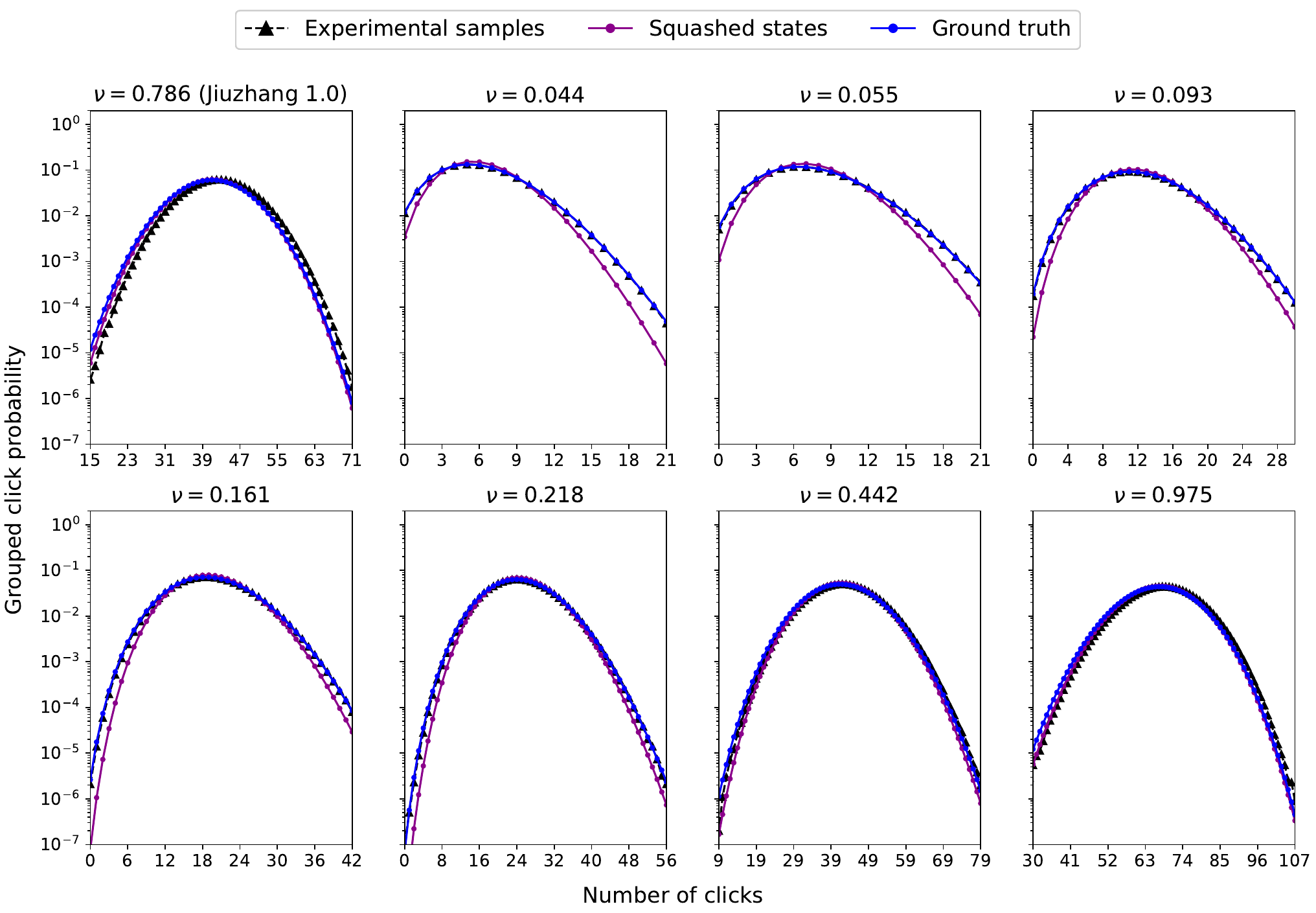}
		\caption{Experimental (black), and estimated squashed states (violet) and ground truth (blue) grouped click probability distributions for the different setups of the Jiuzhang experiments. The $y$-axis of the figures is in logarithmic scale. The uncertainty associated to the estimated distributions is too small to be depicted in this figure. Notice that the ground truth distribution overlaps with the experimental distribution for setups in the low photon number ($\nu$) regime. However, for increasing $\nu$, the squashed states distribution becomes closer to the experimental distribution.
		}
		\label{fig:click_probabilities}
	\end{figure*}

    \begin{table*}[!htp]
		\centering
		\resizebox{\textwidth}{!}{
		\begin{tabular}{|c|c|c|c|cc|cc|}
			\hline
			\multirow{2}{*}{Experiment} &
			\multirow{2}{*}{$P(\text{W})$} &
			\multirow{2}{*}{$w(\mu\text{m})$} &
			\multirow{2}{*}{$\nu$} & \multicolumn{2}{c|}{Ground truth} & \multicolumn{2}{c|}{Squashed states} \\ \cline{5-8}
			\rule{0pt}{11pt} & & & & \multicolumn{1}{c|}{$\bar{C}$} & $\sigma(C)$ &\multicolumn{1}{c|}{$\bar{C}$} & $\sigma(C)$ \\ \hline
			Jiuzhang 1.0 & - & - &$0.786$ & \multicolumn{1}{c|}{$41.042 \pm 0.007$} & $6.509 \pm 0.022$ & \multicolumn{1}{c|}{$41.162 \pm 0.007$} & $6.343 \pm 0.022$ \\ \hline
			\multirow{7}{*}{Jiuzhang 2.0} & 0.5 & 125 & $0.055$ & \multicolumn{1}{c|}{$7.272 \pm 0.002$} & $3.393 \pm 0.002$ & \multicolumn{1}{c|}{$7.378 \pm 0.001$} & $2.91 \pm 0.001$ \\
			\cline{2-8} & 1.412 & 125 & $0.161$ & \multicolumn{1}{c|}{$19.255 \pm 0.003$ } & $5.596 \pm 0.005$ & \multicolumn{1}{c|}{$19.471 \pm 0.003$} & $5.039 \pm 0.005$\\
			\cline{2-8} & 0.15 & 65 & $0.044$ & \multicolumn{1}{c|}{$5.975 \pm 0.001$} & $3.017 \pm 0.001$ & \multicolumn{1}{c|}{$6.057 \pm 0.001$} & $2.616 \pm 0.001$ \\ 
			\cline{2-8} & 0.3 & 65  & $0.093$ & \multicolumn{1}{c|}{$11.94 \pm 0.002$} & $4.325 \pm 0.002$ & \multicolumn{1}{c|}{$12.086 \pm 0.002$} & $3.849 \pm 0.002$\\
			\cline{2-8} & 0.6 & 65  & $0.218$ & \multicolumn{1}{c|}{$24.664 \pm 0.004$} & $6.269 \pm 0.007$ & \multicolumn{1}{c|}{$24.897 \pm 0.004$} & $5.787 \pm 0.007$ \\
			\cline{2-8} & 1.0 & 65  & $0.442$ & \multicolumn{1}{c|}{$41.794 \pm 0.007$} & $7.948 \pm 0.017$ & \multicolumn{1}{c|}{$42.067 \pm 0.007$}  & $7.551 \pm 0.018$ \\
			\cline{2-8} & 1.65 & 65  & $0.975$ & \multicolumn{1}{c|}{$66.866 \pm 0.012$} & $9.018 \pm 0.043$ & \multicolumn{1}{c|}{$67.103 \pm 0.012$} & $8.761 \pm 0.043$\\
			\hline
		\end{tabular}
		}
		\caption{Estimated mean number of clicks, $\bar{C}$, and standard deviation, $\sigma(C)$, for the ground truth and squashed states distributions of the different setups of the Jiuzhang experiments. These results are in excellent agreement with the values shown in Table~\ref{tab:experiment_specifics}.}
		\label{tab:estimations}
	\end{table*}
	
	To take into account that the input states of the Jiuzhang experiments are two-mode squeezed states (TMSS), we modify the transformation matrix that represents the action of the interferometer. As was discussed in Sec.~\ref{sec:distributions}, the input TMSS can be generated by sending SMSS into $50:50$ beamsplitters, whose action is represented by a $K\times K$ matrix $\bm{B}$. Following this idea, we apply the transformation induced by the beamsplitters before applying the transformation of the interferometer over the amplitudes $\bm{\alpha}_k$ and $\bm{\beta}_k$ of the SMSS. Then, we can write the transformed amplitudes $\bm{\bar{\alpha}}_k$ and $\bm{\bar{\beta}}_k$ as $\bm{\bar{\alpha}}_k=\bm{T'}\bm{\alpha}_k$ and $\bm{\bar{\beta}}_k=(\bm{T'})^{\bm{*}}\bm{\beta}_k$, where $\bm{T'}=\bm{T}\bm{B}$.

	This procedure can be readily modified for the case of input squashed states, we need only notice that for these states $\bar{n}_k=\bar{m}_k=\sinh^
	2(r_k)$.

	To assign an uncertainty to the estimation of $\mathrm{Pr}(C)$, we can sort the $\mathcal{N}$ pairs of vectors $\bm{\alpha}_k$ and $\bm{\beta}_k$ in $\mathcal{G}$ groups, each one with $\mathcal{M}=\mathcal{N}/\mathcal{G}$ elements. We then compute $\mathrm{Pr}_{\mathcal{G}}(C)=\frac{1}{\mathcal{M}}\sum_{k=1}^{\mathcal{M}}\tilde{G}(\bm{\bar{\alpha}}_k, \bm{\bar{\beta}}_k;C)$ for each group. Notice that the average of the $\mathrm{Pr}_{\mathcal{G}}(C)$ over all groups is equal to $\mathrm{Pr}(C)$. The uncertainty of the estimation of $\mathrm{Pr}(C)$ corresponds to the standard deviation of the $\mathrm{Pr}_{\mathcal{G}}(C)$ over all groups.
	
	For the estimation of the ground truth and squashed states grouped click probability distributions, $\mathrm{Pr}_{(\text{SQUE})}(C)$ and $\mathrm{Pr}_{(\text{SQUA})}(C)$, we used $\mathcal{N}=10^8$ samples $\bm{\alpha}_k$ and $\bm{\beta}_k$. For the computation of the uncertainty, we sorted these samples in $\mathcal{G}=100$ groups. In Fig.~\ref{fig:click_probabilities} we show the obtained distributions for the different setups of the Jiuzhang experiments. In Table~\ref{tab:estimations} we show the estimated mean number of clicks, $\bar{C}$, and standard deviation $\sigma(C)$ of these distributions using the grouped probabilities. They are in excellent agreement with the quantities reported in  Table~\ref{tab:experiment_specifics} obtained from analytical results.
	
	The functions used for the computation of the grouped click probabilities are included in the library \texttt{thewalrus}~\cite{gupt2019thewalrus} from version 0.20.0 onward.

	\section{\label{app:squashed}Some properties of squashed states}
 
	In this appendix we list some of the most important properties of the squashed states. 

    As was mentioned in Sec.~\ref{sec:distributions}, when losses are incorporated in the input states, squeezed states become squeezed thermal states. The covariance matrix of a single mode squeezed thermal state can be written in the form~\cite{qi2020regimes}
    \begin{equation}
		\bm{\sigma}_{\mathrm{(SQTH)}}=\frac{\hbar}{2}
		\begin{pmatrix}
			1 + \eta(e^{-2r}-1)&0\\
			0&1 + \eta(e^{2r}+1)
		\end{pmatrix},
		\label{eq:sm_sqz_thermal}    
	\end{equation}
    where $r$ is the squeezing parameter, and $\eta>0$ is a transmission rate parameter representing the losses. Notice that $1 + \eta(e^{-2r}-1)<1$ for all $r$, while $1 + \eta(e^{2r}-1)>1$. This indicates that squeezed thermal states still have a quadrature with lower noise than the vacuum. Moreover, it implies that $\bm{\sigma}_{\mathrm{(SQTH)}}\ngeq\frac{\hbar}{2}\mathbb{I}_2$, so these Gaussian states cannot be written as a classical mixture of coherence states (i.e. they are quantum Gaussian states).

    Squashed states, on the other hand, are defined to have the variance of one of their quadratures equal to those of the vacuum, while the remaining quadrature has excess noise proportional to the mean number of photons $\bar{n}$ of the state. The covariance matrix of a single mode squashed state can be written as
	\begin{equation}
		\bm{\sigma}_{\mathrm{(SQUA)}}=\frac{\hbar}{2}
		\begin{pmatrix}
			1&0\\
			0&1+4\bar{n}
		\end{pmatrix}.
		\label{eq:sm_squashed}    
	\end{equation}
	Notice that $\bm{\sigma}_{\mathrm{(SQUA)}}\geq\frac{\hbar}{2}\mathbb{I}_2$, making these states classical mixtures of coherent states (i.e. they are classical Gaussian states). It is possible to show~\cite{qi2020regimes} that the squashed states are the classical Gaussian states with the closest fidelity to squeezed thermal states, making them great candidates to model the effect of losses in the input states of GBS experiments.
 
	Using the methods of Refs.~\cite{yao2022recursive, quesada2019simulating}, it can be shown that the photon number distribution of a Gaussian state with covariance matrix $\bm{\sigma}_{(\mathrm{SQUA})}$ is given by 
	
	\begin{align}
	\begin{split}
		\textrm{Pr}_{\mathrm{(SQUA)}}(N) &= \sqrt{\frac{1}{1+2\bar{n}}}\\
		&\times\left(1-\frac{1}{1+2\bar{n}}\right)^{N}\frac{(2N)!}{(2^N N!)^2}.
		\label{eq:squashed_number_dist} 
	\end{split}
	\end{align}
	This distribution is equivalent to a negative binomial distribution $\textrm{NB}(r, p)$ with $r=1/2$ and $p = 1/(1+2\bar{n})$. With this relation in mind, it can be verified that 
	\begin{equation}
	    \textrm{Var}_{\mathrm{(SQUA)}}(N)=\frac{r(1-p)}{p^2} = \bar{n}(2\bar{n}+1).
	    \label{eq:var_squashed_dist}
	\end{equation}
	
	Moreover, we can compute the second order correlation function, $g^{(2)}_{\mathrm{(SQUA)}}(0)$ as
	\begin{equation}
	    g^{(2)}_{\mathrm{(SQUA)}}(0)=\frac{\langle N^2\rangle-\langle N\rangle}{\langle N\rangle^2} = 3,
	    \label{eq:corr_squashed_dist}
	\end{equation}
	where $\langle\cdot\rangle$ indicates a expected value with respect to distribution $\textrm{Pr}_{\mathrm{(SQUA)}}(N)$. Notice that $\langle N\rangle = \bar{n}$. 
	
	For comparison, recall that the values of $g^{(2)}(0)$ for thermal, coherent and squeezed states are $g^{(2)}_{(\text{THM})}(0) = 2$, $g^{(2)}_{(\text{COH})}(0) = 1$, and $g^{(2)}_{(\text{SQUE})}(0) = 3 + 1/\bar{n}$, respectively.
	
	\bibliographystyle{unsrtnat}
	\bibliography{apssamp}
	
\end{document}